\RequirePackage{lineno}
\documentclass[12pt, aps,prd,superscriptaddress,showpacs,preprintnumbers,amsmath,amssymb,nofootinbib]{revtex4-2}
\newcommand{\RN}[1]{%
  \textup{\uppercase\expandafter{\romannumeral#1}}%
}

\usepackage{graphicx} % Include figure files
\usepackage{dcolumn}  % Align table columns on decimal point
\usepackage{epstopdf}
\usepackage{multirow}
\usepackage{color}
\usepackage{comment}
\usepackage[bookmarks]{hyperref}
\usepackage{amsmath}
\usepackage{nccmath}
\usepackage[symbol]{footmisc}
\usepackage{ulem}
\usepackage{orcidlink}

\graphicspath{{ps}}

\newcommand{\Lc}{\Lambda_c^+}
\newcommand{\Lcneutral}{\Lambda_c^+ \to p K_S^0 \pi^0}
\newcommand{\Kshort}{K_S^0}
\newcommand{\Lccharged}{\Lambda_c^+ \to p K^- \pi^+}
\newcommand{\massMeV}{{\rm{MeV}}/c^2}
\newcommand{\momentumMeV}{{\rm{MeV}}/c}
\newcommand{\massGeV}{{\rm{GeV}}/c^2}
\newcommand{\massGeVsq}{{\rm{GeV}^2}/c^4}

\newcommand{\etal}{\it et al}
\begin{document}
%\begin{flushfright}
%Publication number: BELLE2-PUB-DRAFT-2023-018\\
%Authors: YoungJun Kim, Jung Keun Ahn, and Seongbae Yang\\
%Referees: Committee: John Yelton (chair), Masayuki Niiyama, and %Elena Solovieva\\
%Intended journal: {\it PRD}\\
%\end{flushright}
%\vspace*{-100pt}

%\preprint{
%\vbox{ 	\vspace*{-0.6cm}					
%                        \hbox{BELLE2-PUB-DRAFT-2023-018}                    
%                        \hbox{Intended Journal: {\it PRD}}
%                        \hbox{Authors: YoungJun Kim, Jung Keun Ahn, and Seongbae Yang}
%                        \hbox{Committee: John Yelton (chair), Masayuki Niiyama, and Elena Solovieva}
%                        \hbox{Belle II Preprint: 2024-022}                                      \hbox{KEK Preprint: 2024-20}
%    \hbox{Belle Note No.: 1554}
%    \hbox{Draft version: 48, Date: \today}
%}}

\title{ Measurement of the Branching Fraction of $\Lambda_c^+ \to p K_S^0 \pi^0$ at Belle}

%%%% >>>>> insert the authorlist here. BEFORE the abstract !!!!! <<<<<
%%%% >>>>> from the authorship confirmation web page
%%% Name the file author.tex and use \input{author} to insert into your latex file.

%%% Paper:    Lambda_c+ to p KS0 pi0
%%% Journal:  Physical Review D
%%% Contacts: Y.J. Kim, S.B. Yang, J.K. Ahn
%%% ====================================================================
%%% Use \input{pub058-orcid} to insert this material into your latex file.

  \author{I.~Adachi\,\orcidlink{0000-0003-2287-0173}} % 2590
% \author{K.~Adamczyk\,\orcidlink{0000-0001-6208-0876}} % 2239
  \author{L.~Aggarwal\,\orcidlink{0000-0002-0909-7537}} % 10083
% \author{P.~Ahlburg\,\orcidlink{0000-0002-9832-7604}} % 2367
  \author{H.~Ahmed\,\orcidlink{0000-0003-3976-7498}} % 11323
  \author{J.~K.~Ahn\,\orcidlink{0000-0002-5795-2243}} % 7423
  \author{H.~Aihara\,\orcidlink{0000-0002-1907-5964}} % 2223
  \author{N.~Akopov\,\orcidlink{0000-0002-4425-2096}} % 9443
  \author{M.~Alhakami\,\orcidlink{0000-0002-2234-8628}} % 28103
  \author{A.~Aloisio\,\orcidlink{0000-0002-3883-6693}} % 2194
% \author{S.~Al~Said\,\orcidlink{0000-0002-4895-3869}} % 6823
  \author{N.~Althubiti\,\orcidlink{0000-0003-1513-0409}} % 21524
% \author{L.~Andricek\,\orcidlink{0000-0003-1755-4475}} % 2098
  \author{M.~Angelsmark\,\orcidlink{0000-0003-4745-1020}} % 13963
% ...
% 13403
% \author{S.~Zou\,\orcidlink{0000-0003-3377-7222}} % 19363
\collaboration{The Belle and Belle II Collaborations}

\begin{abstract}
We report a precise measurement of the ratio of branching fractions
$\mathcal{B}(\Lcneutral)/\mathcal{B}(\Lccharged)$ 
using 980 fb$^{-1}$ of $e^+e^-$ data 
from the Belle experiment. We obtain a value of $\mathcal{B}(\Lcneutral)/\mathcal{B}(\Lccharged)=0.339\pm 0.002\pm 0.009$,
where the first and second uncertainties are statistical and systematic, respectively.
This Belle result is consistent with the previous measurement from the CLEO experiment but has a fivefold improvement in precision.
By combining our result with the world average $\mathcal{B}(\Lccharged)$, we obtain the absolute branching fraction 
$\mathcal{B}(\Lcneutral)=(2.12\pm 0.01\pm 0.05 \pm 0.10)\%$, where the uncertainties are statistical, systematic, and the uncertainty in the absolute branching fraction scale $\mathcal{B}(\Lccharged)$, respectively. This measurement can shed light on hadronic decay mechanisms in charmed baryon decays.

\end{abstract}

\maketitle

\tighten

\section{Introduction}

\begin{figure}[!htb]
\centering
  \includegraphics[width=0.3\textwidth]{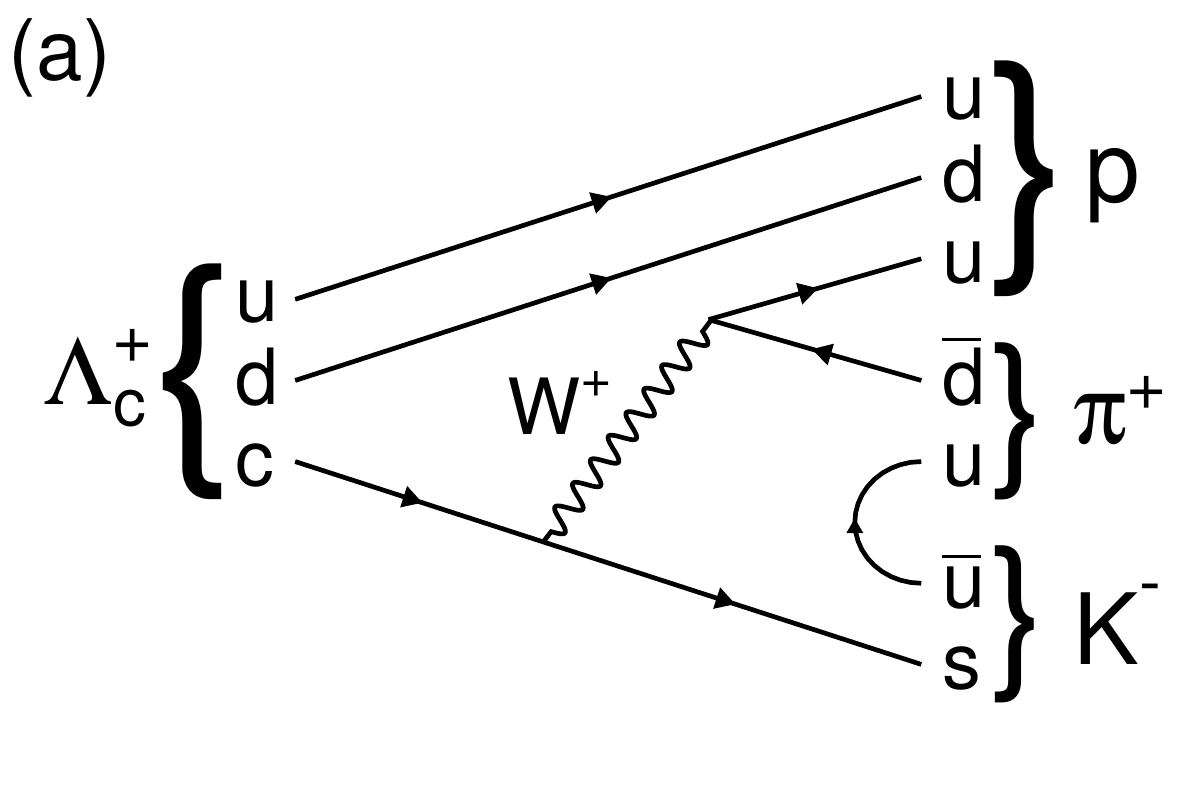}
  \includegraphics[width=0.3\textwidth]{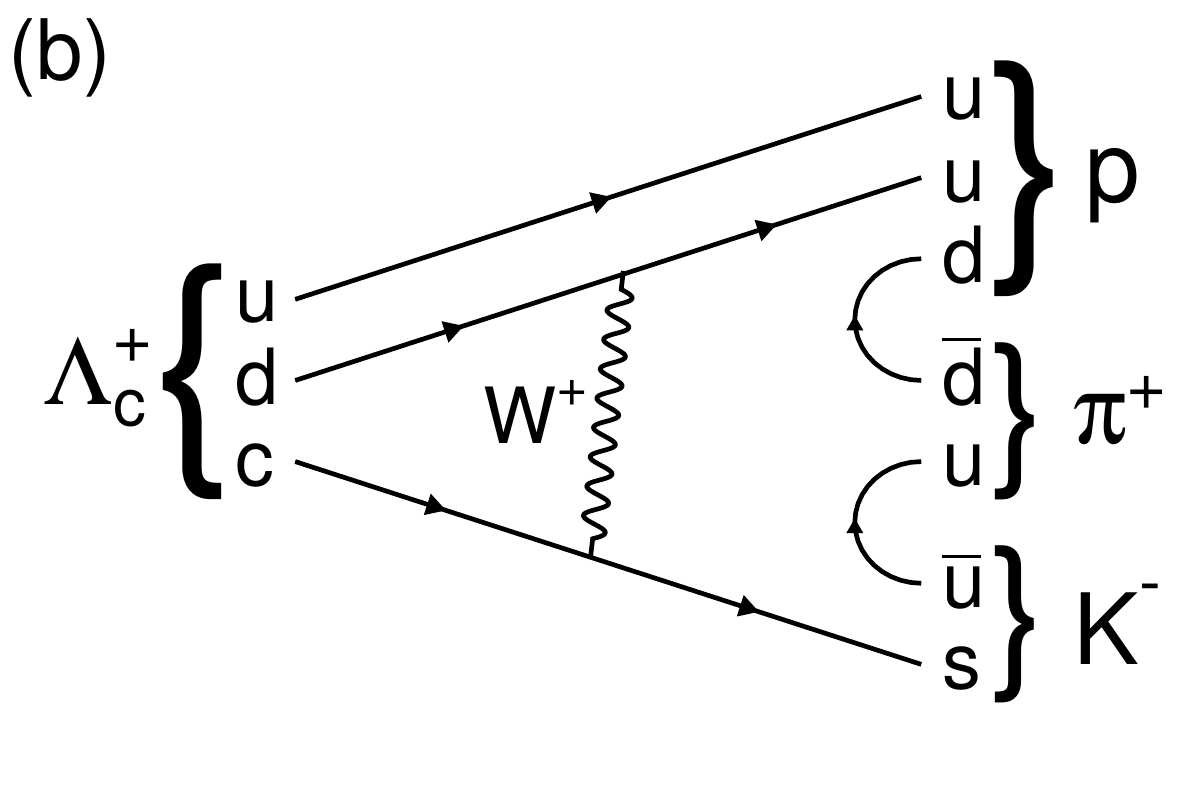}
  \includegraphics[width=0.3\textwidth]{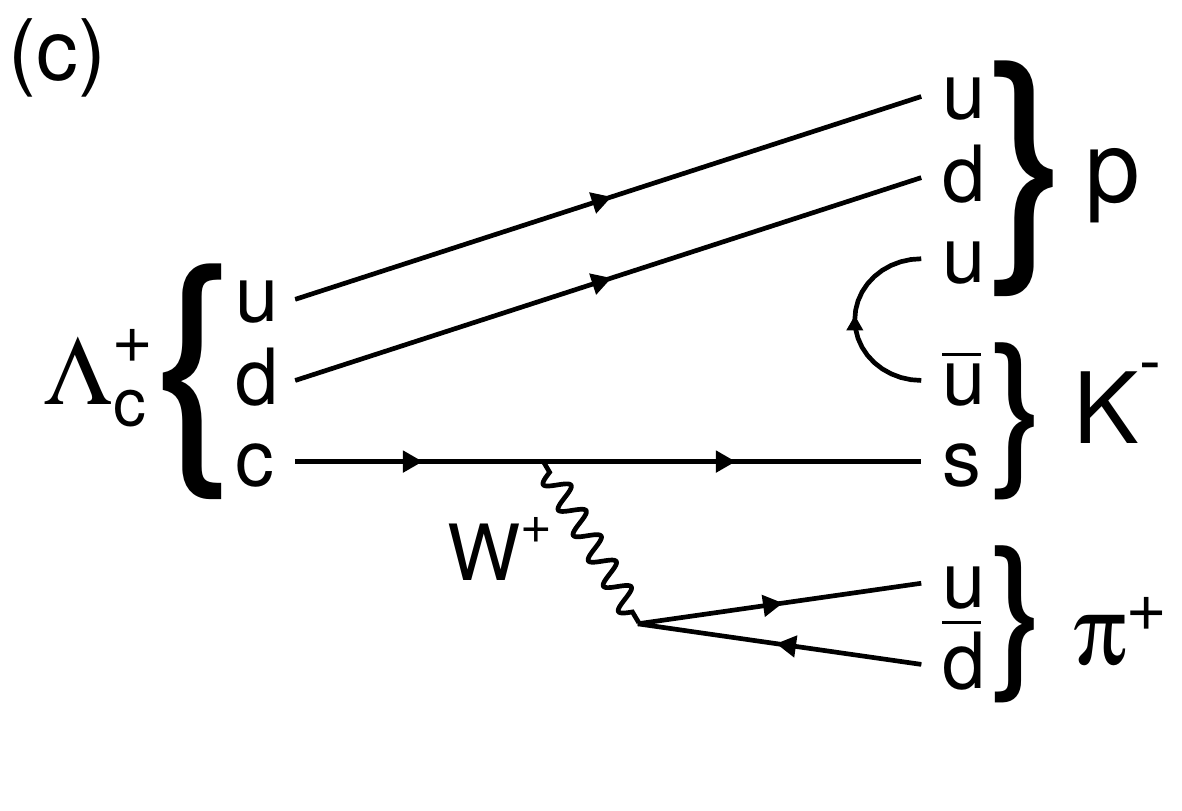}
  \raggedright
  \includegraphics[width=0.3\textwidth]{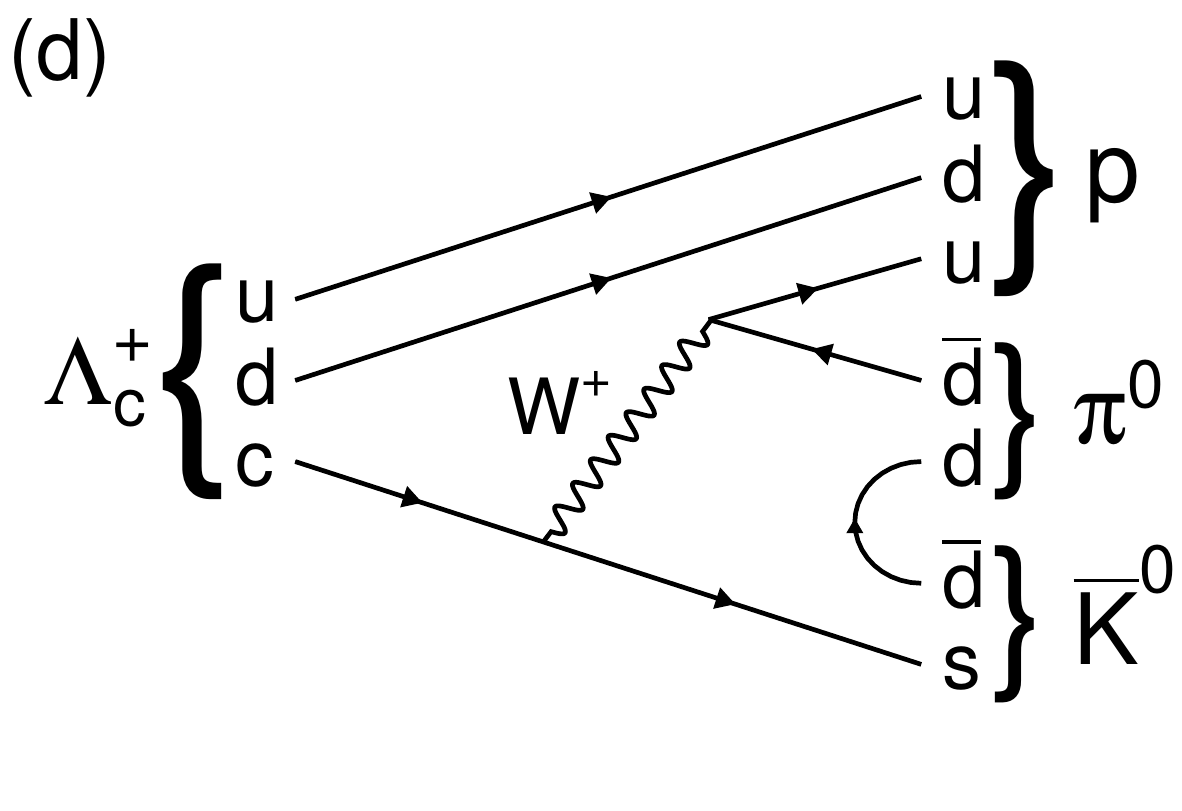}
  \includegraphics[width=0.3\textwidth]{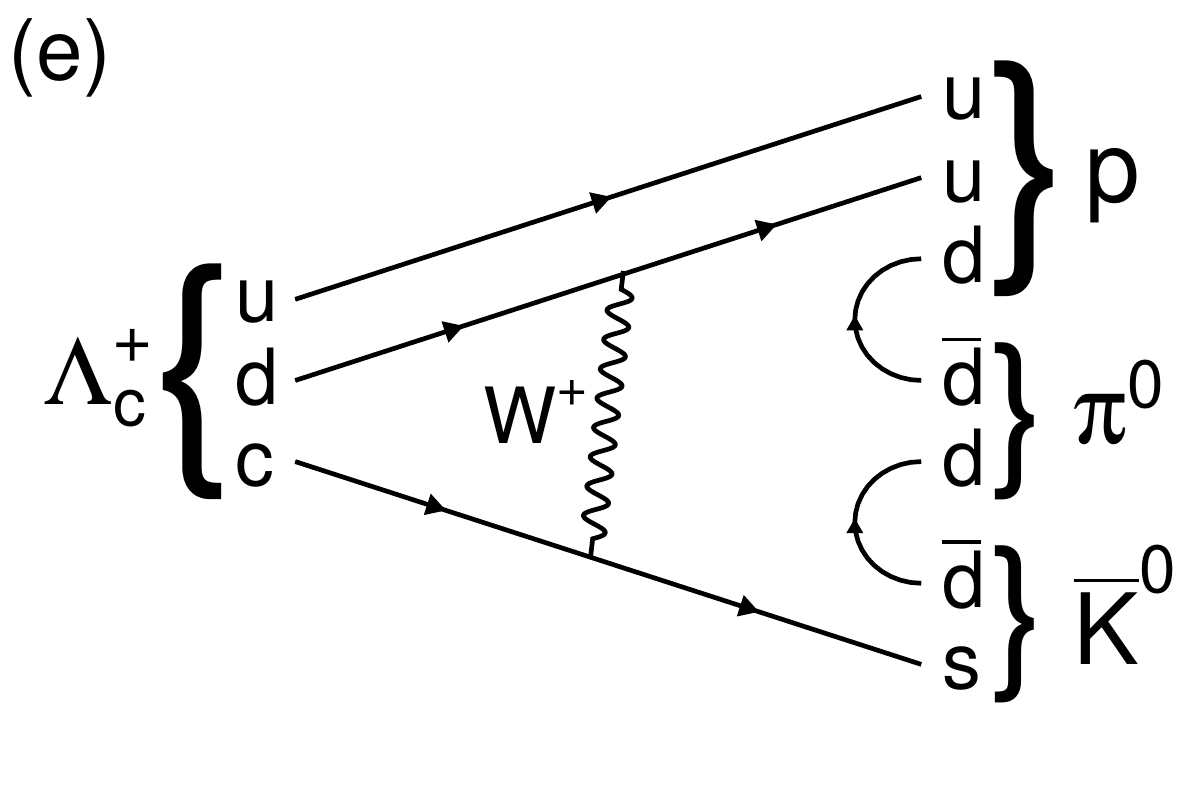}
\caption{
Typical Feynman diagrams for internal $W$ emission (a), internal $W$ exchange (b), and external $W$ emission (c) processes in $\Lccharged$ decays, 
and internal $W$ emission (d) and internal $W$ exchange (e) processes in $\Lcneutral$ decays.
}
\label{fig:feynman}
\end{figure}

The nonleptonic weak decays of $\Lambda_{c}^{+}$ provide a
unique testing ground for understanding the factorization scheme 
involving the $c\to s$ transition and the final-state interactions.
Among the possible final states, $N\bar{K}\pi$ decays are particularly 
useful for examining the isospin properties of the weak interaction in $\Lambda_{c}^{+}$ decay~\cite{conjugate}. 
The $\Delta S=1$ Cabibbo-allowed transition is governed by $c\to su\bar{d}$, 
so iso-singlet $\Lambda_c^+$ decays result in a final state with $I=I_3=1$. 
In the $\Lambda_c^+\to N\bar{K}\pi$ decays, the $N\bar{K}$ state can have isospin 0 or 1. 
Thus, the sum of isospin amplitudes of the three decay modes,
$\sqrt{2}\mathcal{A}(p\bar{K}^{0}\pi^{0}) + \mathcal{A}(pK^{-}\pi^{+}) 
+ \mathcal{A}(n\bar{K}^{0}\pi^{+})$,
is zero according to isospin symmetry~\cite{Lu:2016ogy, Ablikim:2017, Gronau}, which imposes useful constraints on the branching ratios of the nonleptonic decay channels.
In the quark-diagram schemes for $\Lambda_c^+\to N\bar{K}\pi$ decays, as shown in Fig.~\ref{fig:feynman}, direct $\pi^+$ emission can involve a color-allowed
factorizable process with external $W^+$ emission~(Fig.~\ref{fig:feynman}(c)) but a $\pi^0$ cannot be produced in this process. The dominant
contributions in the $N\bar{K}\pi^0$ decays instead are from color-suppressed internal $W^+$ emission and
internal flavor conversion involving the subprocess $cd\to su$ with $W^+$ exchange.

Since direct $\pi^+$ emission leaves the $ud$ diquark in the 
iso-singlet $\Lambda_c^+$ as a spectator, the $sud$ cluster in the final state is a pure $I=0$ state, 
thus favoring the $I=0$ $\bar{K}N$ state~\cite{Miyahara:2015, Ahn:2019}. 
Hence, if the direct $\pi^+$ emission process is dominant, the two-body decay $\Lc \to \Lambda \pi^+$ should be greatly favored over $\Lc \to \Sigma^0 \pi^+$. 
Moreover, as the $\Lambda_c^+$ contains an anti-symmetric quark pair $(ud-du)c$, factorizable processes are suppressed for $\Lambda_c^+$ decays involving a baryon decuplet with totally flavor symmetric quark content~\cite{hsiao}.
However, the experimentally determined branching fractions for the two decays are
comparable, indicating that the contributions from color-suppressed $W^+$
emission and $W^+$ exchange processes are large.
The rescattering of the $N\pi$ pair in the final state can populate $N(1535)$ and $N(1650)$ resonances,
and the $N\bar{K}$ final-state interaction generates $\Lambda(1405)$ and $\Lambda(1670)$ states~\cite{pavao}.

An observation of a narrow structure at the $\Lambda\eta$ threshold
in the $K^-p$ invariant mass spectrum 
in $\Lambda_c^+\to pK^-\pi^{+}$ reported by the Belle collaboration~\cite{Yang:2015ytm} has attracted significant attention. A recent analysis, attempting to shed light on the nature of the structure, indicates a $\Lambda\eta$ cusp effect, enhanced by the $\Lambda(1670)$ pole~\cite{Belle:2022cbs}.
Another study suggests that the effect giving rise to the narrow structure is enhanced by a triangle singularity involving rescattering via the near-lying $\Lambda a_0(980)^+$ or $\eta\Sigma(1660)^+$ scattering~\cite{Liu}.
In $\Lccharged$ process, isospin symmetry implies that the partial branching ratio of $\Lambda_c^+\to\Sigma^{\ast0}\pi^+$ equals that of 
$\Lambda_c^+\to\Sigma^{\ast+}\pi^0$, whereas the $\Lambda_c^+\to\Delta^{++}K^-$ decay is three times larger than the branching fraction of $\Lambda_c^+\to\Delta^+\bar{K}^0$~\cite{hsiao, savage}.
In the $\Lambda_{c}^{+}\to p\bar{K}^{0}\pi^{0}$ decay, only $\Sigma^{\ast+}$ resonances are possible in the $N\bar{K}$ system.
Therefore, a precise measurement of the relative branching fraction 
for $\Lcneutral$ as well as the investigation of intermediate resonances provides stringent tests of isospin symmetry and could help to better understand non-factorizable
processes in the non-leptonic decay of charmed baryons.

The absolute branching fractions for $\Lambda_c^+\to n\bar{K}^{0}\pi^{+}$  and  $\Lcneutral$ decays
are reported from BESIII to be $\mathcal{B}(\Lambda_{c}^{+} \to n K^{0}_{S} \pi^{+}) = (1.82 \pm 0.25)\%$ and  $\mathcal{B}(\Lcneutral) = (1.87 \pm 0.14)\%$, respectively~\cite{Ablikim:2017}. 
The branching fraction of $\Lcneutral$ 
relative to $\Lccharged$ reported from CLEO is $0.33 \pm 0.05$~\cite{CLEO:1998dce}. 
Here we report a precise measurement of the relative branching fraction of 
$\Lcneutral$ compared with $\Lccharged$ using Belle data. In addition, we present the first investigation of the intermediate resonances in $\Lcneutral$ decays.

\section{The Data Sample and the Belle Detector}

The branching fractions are measured based on a data sample 
obtained at or near $\Upsilon(1S)$, 
$\Upsilon(2S)$, $\Upsilon(3S)$, $\Upsilon(4S)$ and $\Upsilon(5S)$
with the Belle detector at the KEKB asymmetric energy $e^+e^-$ collider~\cite{Kurokawa:2001nw}. 
The full Belle data sample has an integrated luminosity of 980 $\rm{fb}^{-1}$. 
The Belle detector was a large-solid-angle magnetic spectrometer 
comprising a silicon vertex detector~(SVD), a central drift chamber~(CDC), 
an array of aerogel threshold Cherenkov counters~(ACC), 
a barrel-like arrangement of time-of-flight scintillation counters~(TOF), 
and an electromagnetic calorimeter comprising CsI(Tl) crystals~(ECL) 
located inside a superconducting solenoid that provided a 1.5 T magnetic field. 
An iron flux return located outside the coil 
was employed to detect $K_L^0$ mesons and identify muons. 
The Belle detector is described in detail in Ref.~\cite{Belle:2000cnh}.

The Monte Carlo~(MC) samples used in the simulation studies are generated using EvtGen~\cite{Lange:2001uf} 
and {\sc PYTHIA}~\cite{Sjostrand:2006za}, 
and propagated through a {\sc GEANT}3 model of the full detector~\cite{Brun}. 
The final-state radiation process is simulated 
using the {\sc PHOTOS}~\cite{Barberio:1993qi} package in EvtGen.
A signal MC sample is generated via $e^+ e^-\to c \bar{c} \to \Lc + X$ 
to study the reconstruction efficiency and signal shape functions.
A Belle generic MC simulated data sample including $\Upsilon(4S)\to B\bar{B}$, $\Upsilon(5S)\to B^{(*)}_{(s)}\bar{B}^{(*)}_{(s)}$, $\Upsilon(1S,2S,3S)$ decays and $e^+e^-\to q\bar{q}~(q=u,d,s,c)$ with the same integrated luminosity as the data is used to optimize the selection criteria.

\section{Event Selection}

We reconstruct $\Lcneutral$ with $\Kshort \to \pi^+ \pi^-$ and $\pi^0 \to \gamma\gamma$.
The event selection criteria are optimized using a generic MC sample, with a figure-of-merit~(FoM) defined as $N_{\rm{sig}}/\sqrt{N_{\rm{sig}} + N_{\rm{bkg}}}$, 
where $N_{\rm{sig}}$ is the number of signal events and $N_{\rm{bkg}}$ is
the number of background events.
The latter are obtained in the $p\Kshort\pi^0$ invariant mass region between 2.263~$\massGeV$ and 2.306~$\massGeV$.
   
The likelihood $\mathcal{L}_i ~(i = \pi^\pm ,~K^\pm,~p^\pm)$ is calculated by combining information from the ACC, CDC, and TOF detectors. 
The likelihood ratio between hypotheses $i$ and $i^\prime$ is defined as $\mathcal{R}(i|i^\prime)
=\mathcal{L}_i/(\mathcal{L}_i+\mathcal{L}_{i^\prime})$.
Charged tracks must satisfy $\mathcal{R}(p|K)> 0.9$ and $\mathcal{R}(p|\pi)> 0.9$ 
to be considered as proton candidates. 
Furthermore, the electron likelihood ratio~($\mathcal{R}(e)$), obtained from
ACC, CDC, and ECL information, should be smaller than 0.9 for the proton candidates.
For all proton candidates, the distance-of-closest-approach~(DOCA) to the beam interaction point~(IP) 
must be smaller than 2.0 cm along the beam direction~($z$) and smaller than 0.1 cm 
in the transverse direction~($r$). Furthermore, at least one hit in SVD is required.
After applying the selection criteria, the PID efficiency for proton candidates is
83\% in the typical momentum range of $\Lc$ decays.

We reconstruct $K_S^0$ candidates using pairs of oppositely charged particles
assumed to be pions. We use a neural network algorithm involving
the $\Kshort$ momentum in the laboratory frame, 
the distance between two charged pion tracks along the $z$ axis,
the flight length of $\Kshort$ projected onto the $r$ plane,
the angle between the $\Kshort$ momentum and the vector 
from IP to $\Kshort$ decay vertex in the laboratory frame, longer and shorter DOCAs in the $r$-direction of charged pions,
the angle between the $\Kshort$ momentum in the laboratory frame 
and the charged pion momentum in the $\Kshort$ rest frame,
the number of CDC hits from each $\pi^{\pm}$ track 
and the presence or absence of SVD hits~\cite{Belle:2018xst}. 
In addition, we perform a mass-constrained fit to the $K_S^0$ candidates in order to improve the $\Kshort$ momentum resolution. The $\chi^2$ value of the mass-constrained vertex fit to the $\pi^+$ and $\pi^-$ tracks with a common vertex is required to be smaller than 40.

ECL clusters that do not have matching tracks in the CDC are identified as photons, and
the $\pi^0$ candidates are reconstructed from photon pairs.
For each photon, the energy deposited in the ECL must exceed 
50~(100)~MeV if the cluster is found in the barrel~(end-cap) region~\cite{Belle:2000cnh}. 
The ratio of energy deposits in the $3 \times 3$ array of 
crystals, centered in the crystal with the highest energy, 
to that of $5 \times 5$ crystal array must exceed 0.9.
We select the $\pi^{0}$ candidates within the $M(\gamma\gamma)$ range
from $120~\massMeV$ to $150~\massMeV$, corresponding to approximately three standard deviations ($\sigma$) in the $M(\gamma\gamma)$ resolution.
The momentum of the $\pi^{0}$ candidate must be greater than
400~$\momentumMeV$ in the laboratory frame.
A mass-constrained fit is also performed on the $\pi^{0}$ candidates to improve their momentum resolution.
%, requiring the $\chi^2$ value of the fit to be smaller than 100.

As a final step,
the proton, $K_{S}^{0}$, and $\pi^{0}$ candidates are combined
to reconstruct $\Lambda_{c}^{+}$ candidates.
The scaled momentum $x_p$ is defined as $x_{p} = p^{\ast}c/\sqrt{s/4-M^2 c^{4}}$, where $p^\ast$ is the momentum of the $\Lc$ candidate in the center-of-mass frame, $s$ is the square of the beam center-of-mass energy and $M$ is invariant mass of the $\Lambda_c^+$ candidate.
The requirement $x_{p}>0.54$ reduces the combinatorial background,
particularly from $B$ meson decays.
We perform a vertex fit to the three decay products requiring the reconstructed $K_S^0$ and $\pi^0$ originate from the $\Lc$ decay vertex.
The $\chi^2$ value of the vertex fit is required to be smaller than 40.

For $\Lccharged$ decays, we reconstruct $\Lambda_{c}^{+}$ candidates
using the event selection criteria typically used in other $\Lambda_c^+$ analyses with Belle~\cite{Yang:2015ytm}, except that the $x_{p}$ cut-off value
is the same as in $\Lcneutral$ decays.
For proton candidates, the same selection criteria are used as in our signal mode.
For $K^{-}$ and $\pi^{+}$ candidates, the requirements on $\mathcal{R}(e)$, DOCAs in $z$- and $r$-directions, and SVD hits are identical to those for proton candidates.
However, the PID requirements are $\mathcal{R}(K|\pi)>0.9$
and $\mathcal{R}(K|p)>0.4$ for $K^{-}$
and $\mathcal{R}(\pi|K)>0.4$ and
$\mathcal{R}(\pi|p)>0.4$ for $\pi^{+}$. The PID efficiencies of $K$ and $\pi$ are 82\% and 94\%, respectively, in the typical momentum range of the decays.
We fit the three decay products to a common vertex. The $\chi^2$ value of the vertex fit is required to be smaller than 40.

After applying all the selection criteria to the data, we observe an average of 1.04 and 1.02 candidates per event for the $\Lcneutral$ and $\Lccharged$ modes within the invariant mass ranges $2.263~\massGeV < M(pK_{S}^0\pi^0) < 2.306~\massGeV $
and $2.274~\massGeV < M(pK^-\pi^+) < 2.298~\massGeV $, respectively.
In addition, we find that approximately 4.0\% and 1.8\% of events in these modes contain multiple signal candidates. Since these multiple candidates do not contribute to the peaking background in a study of the generic MC simulation sample, we retain all candidates for further analysis.

%%%%%
\begin{figure}[!htb]
\centering
  \includegraphics[width=0.65\textwidth]{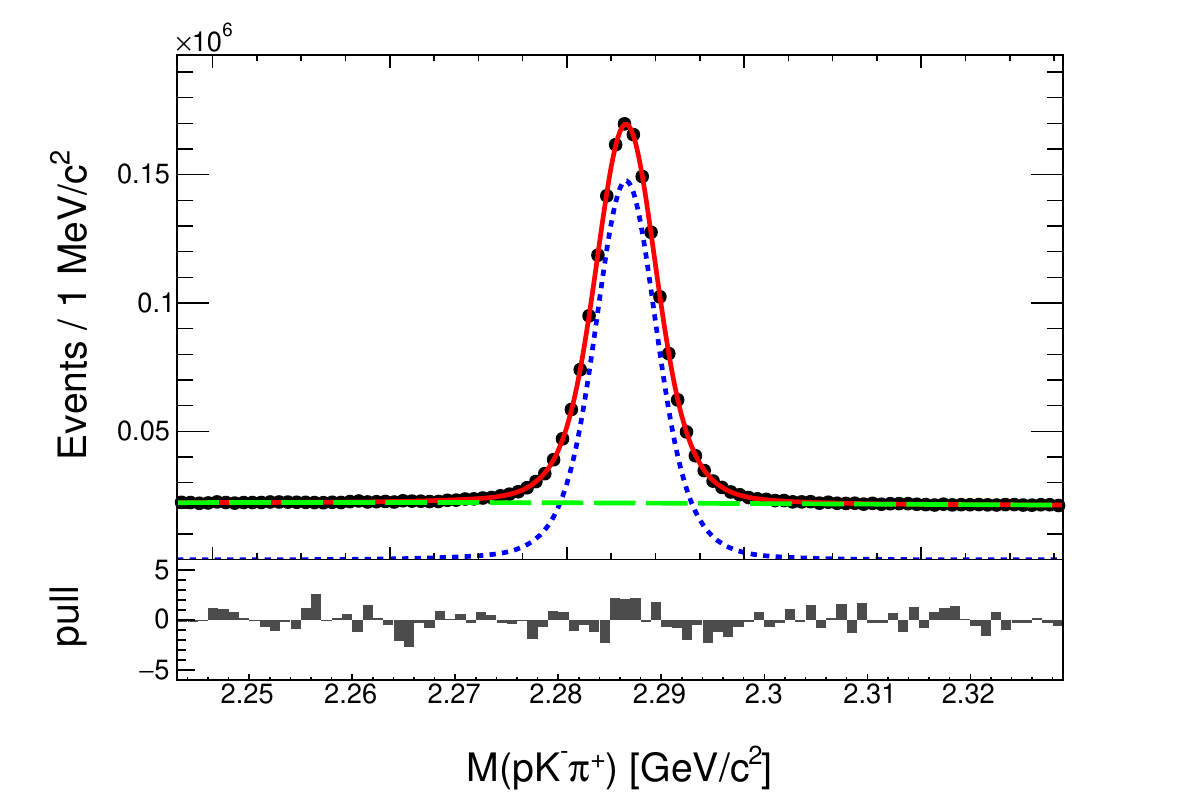}
  \includegraphics[width=0.65\textwidth]{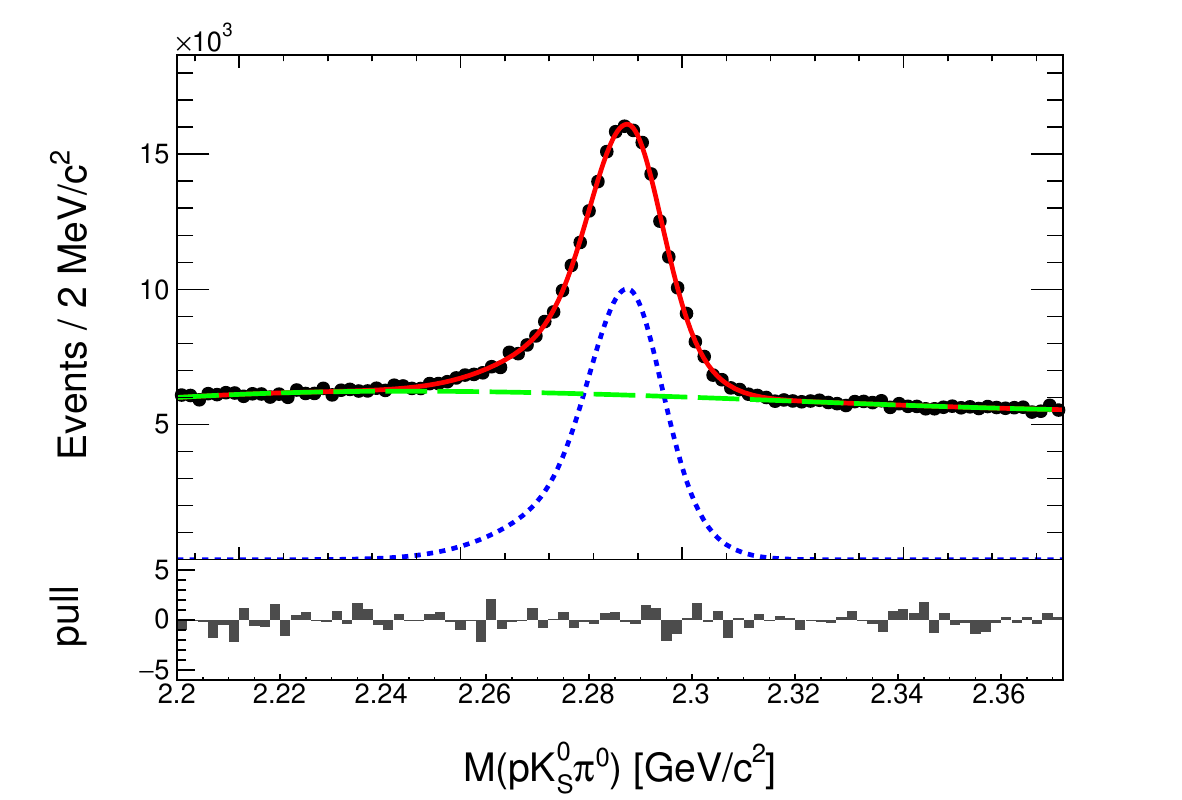}

\caption{Invariant mass distributions of $\Lc$ candidates and fit results for $\Lccharged$~(top) and $\Lcneutral$~(bottom). The total fit, signal, and background are shown by solid red, dashed blue, and long dashed green curves, respectively.}
\label{fig:Lcmass}
\end{figure}
%%%%%%

\section{Signal Extraction and Efficiency Correction}

Figure~\ref{fig:Lcmass} shows the $M(pK^-\pi^+)$ and $M(pK^0_S\pi^0)$ distributions after 
applying the event selection method described in the previous section. 
We perform a binned extended maximum likelihood fit to extract the signal decay yields from the invariant mass distributions.
The $\Lambda^+_c$ signal peak in the $M(pK^-\pi^+)$ distribution, which corresponds to mass resolution due to detector response, is parameterized by a sum of two Gaussian functions and one bifurcated Gaussian function, with each function sharing a common mean.
To accurately model the energy loss associated with $\pi^0$ daughter $\gamma$'s in the $\Lcneutral$ decay, its signal function is taken as the sum of two bifurcated Gaussian functions sharing a common mean. 
A third-order polynomial function represents
the combinatorial backgrounds for $M(pK^-\pi^+)$ and $M(pK^0_S\pi^0)$ fits. 
The extracted $\Lambda^+_c$ yields for $\Lccharged$ and $\Lcneutral$ decays 
are $(1.405 \pm 0.003)\times 10^6$ and $(1.283 \pm 0.010)\times 10^5$, respectively, 
where the uncertainties are purely statistical. \par

The mass resolution parameters obtained by the fit for $\Lccharged$ decay are as follows:
two Gaussians with $\sigma$ values of $(3.17 \pm 0.02)~\massMeV$ and $(4.80 \pm 0.07)~\massMeV$ respectively, and a bifurcated Gaussian with $\sigma_{\rm{left}}= (19.5 \pm  0.39)~\massMeV$ and $\sigma_{\rm{right}} = (10.9 \pm 0.32)~\massMeV$. The yield fractions for the two Gaussians are 
$(50.2 \pm 1.3)\%$ and $(42.3\pm 0.6)\%$. For $\Lcneutral$ decay, there are two bifurcated Gaussians with $\sigma_{\rm{left}}$ and $\sigma_{\rm{right}}$ as follows: $(8.19 \pm 0.23)~\massMeV$ and $(7.31 \pm 0.27)~\massMeV$ for the first Gaussian, and $(19.5 \pm  0.22)~\massMeV$ and $(11.6 \pm 0.30)~\massMeV$ for the second Gaussian. The yield ratio of the first Gaussian to the second is $1.07 \pm 0.07$. The uncertainties of these values are statistical only. 

%%%%%%
\begin{figure}[!htb]
\centering
  \includegraphics[width=0.48\textwidth]{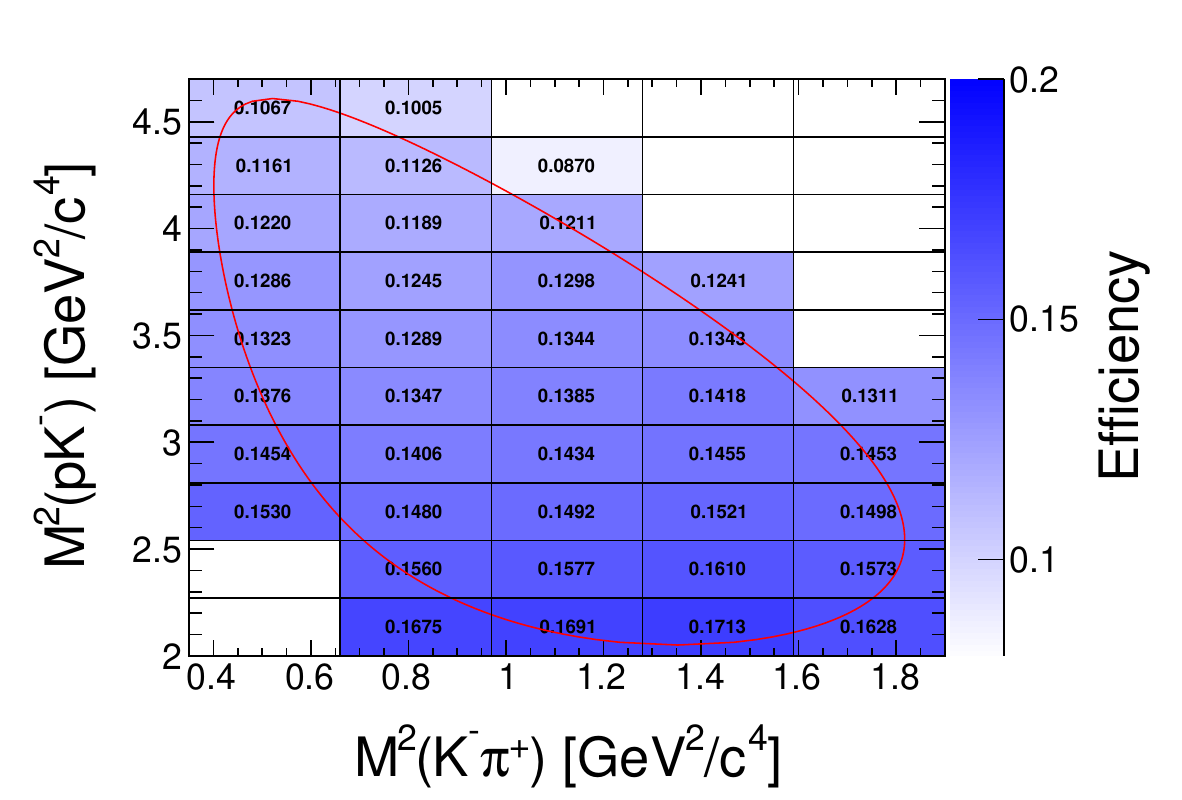}
  \includegraphics[width=0.48\textwidth]{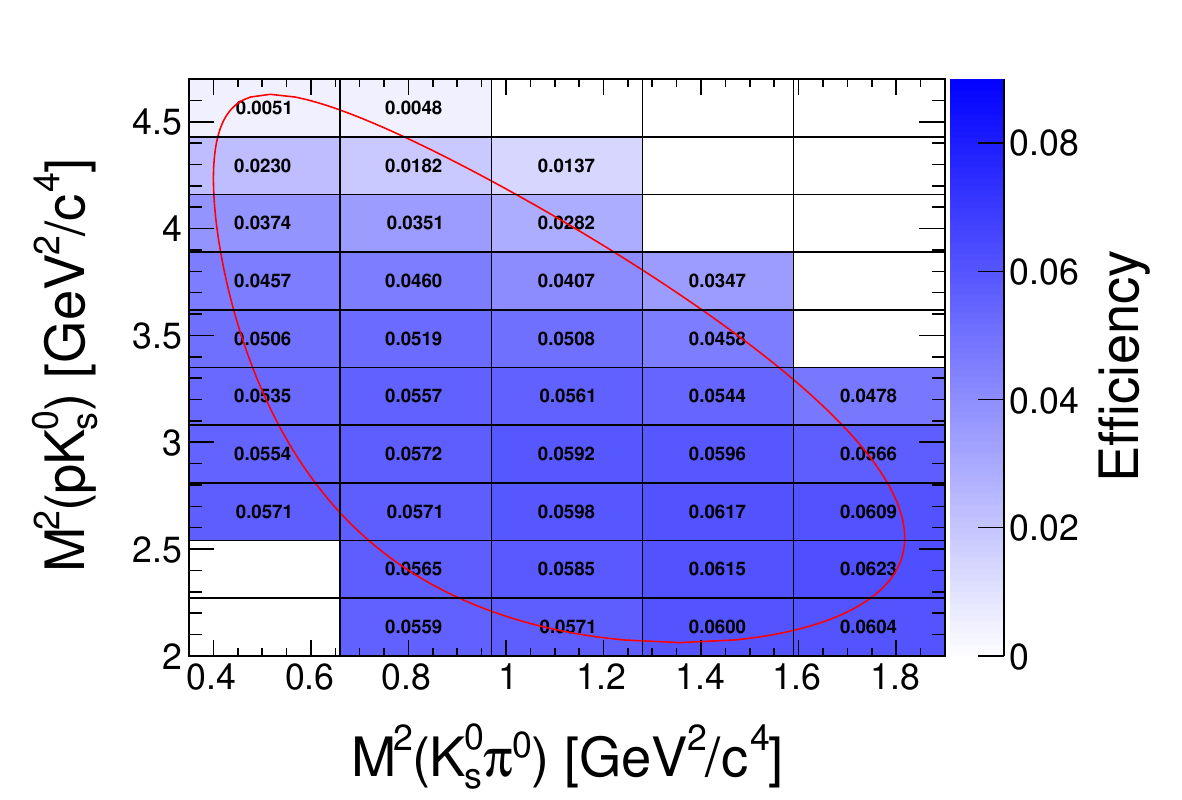}
\caption{Distributions of average reconstruction efficiencies over the Dalitz plots divided into the 10~$\times$~5 bins of $M^2(pK)$ vs. $M^2(K\pi)$ for the $\Lccharged$~(left) and the $\Lcneutral$~(right). Red contours represent the kinematic boundaries of the Dalitz plot, assuming the nominal $\Lambda_c^+$ mass.}
\label{fig:efficiency}
\end{figure}

%%%%%

%%%%%%
\begin{figure}[!htb]
\centering
  \includegraphics[width=0.48\textwidth]{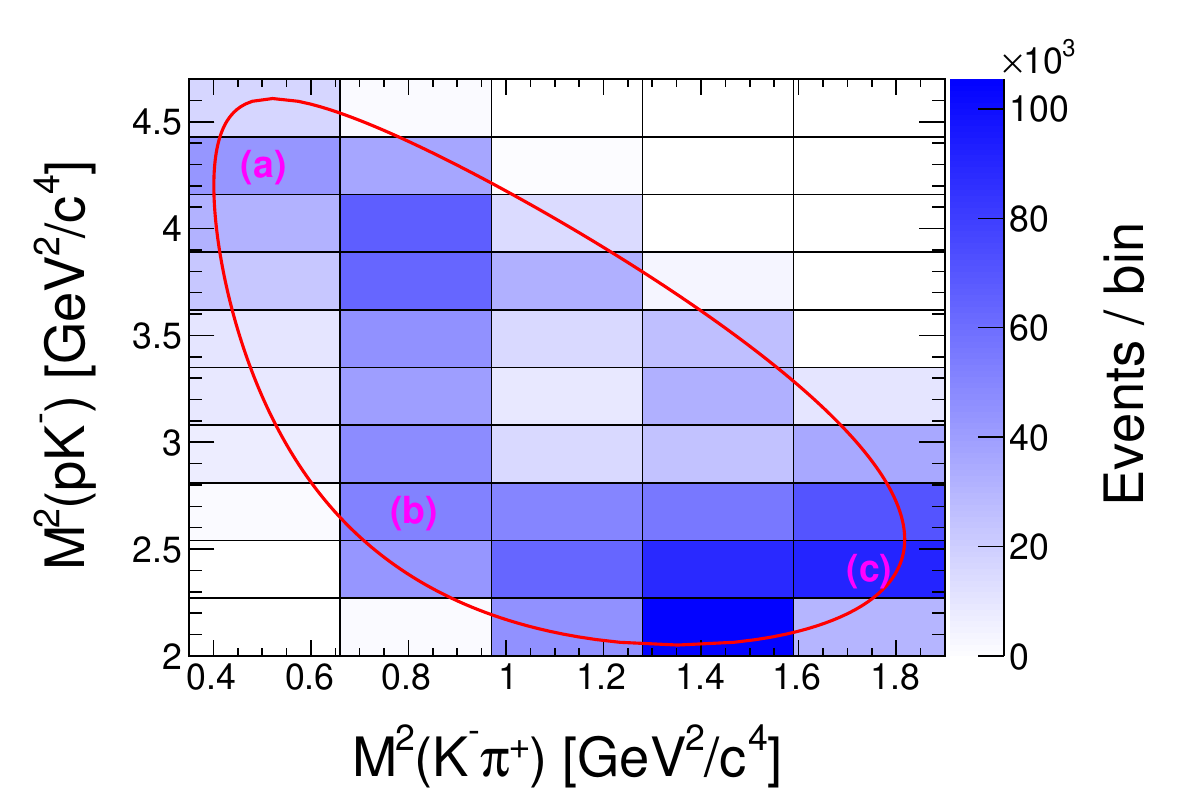}
  \includegraphics[width=0.48\textwidth]{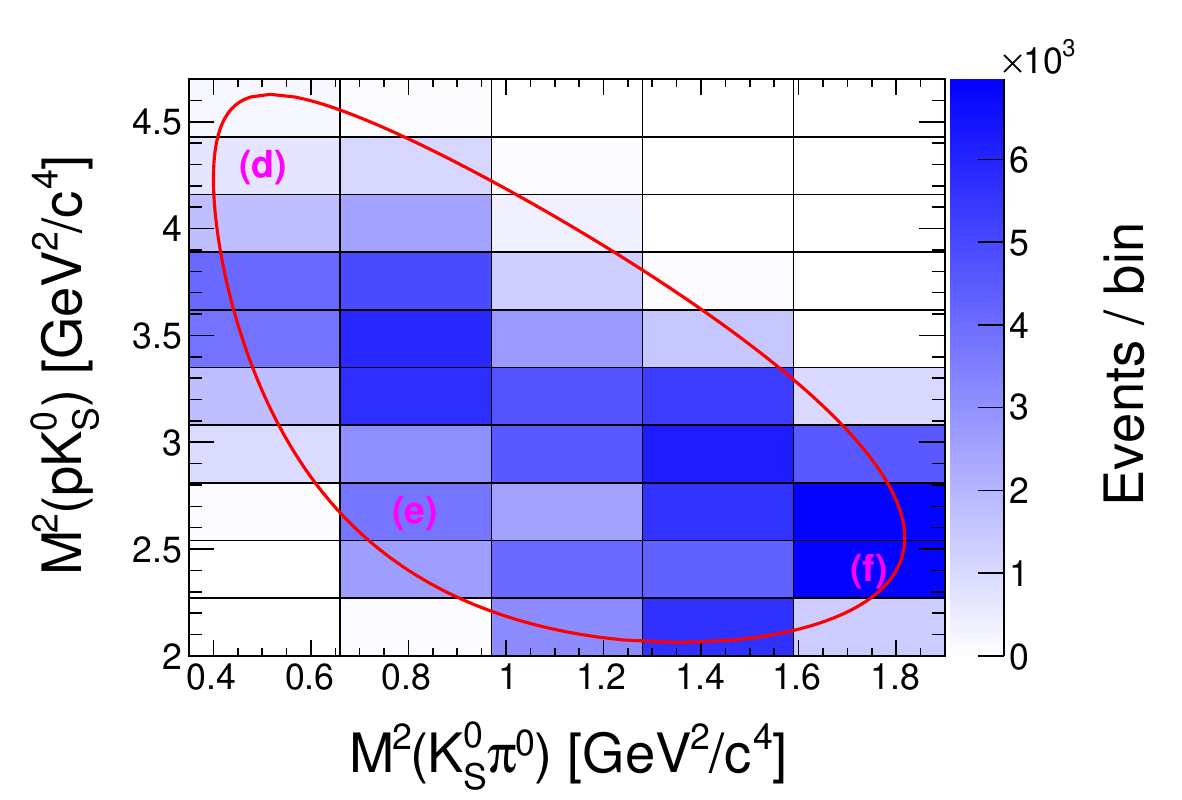}
\caption{Extracted signal yield for each Dalitz bin for 
 $\Lambda_{c}^{+} \rightarrow pK^{-}\pi^{+}$~(left) and $\Lambda_{c}^{+} \rightarrow pK_{S}^{0}\pi^{0}$~(right). Fit results in the sample bins labeled from~(a) to~(f) are shown in Fig.~\ref{fig:exampleFit}.
Red contours represent the Dalitz plot boundaries assuming the nominal $\Lambda_c^+$ mass.}
\label{fig:yield}
\end{figure}
%%%%%%
%%%%%%%%
\begin{figure}[!htb]
\centering
  \includegraphics[width=0.95\textwidth]{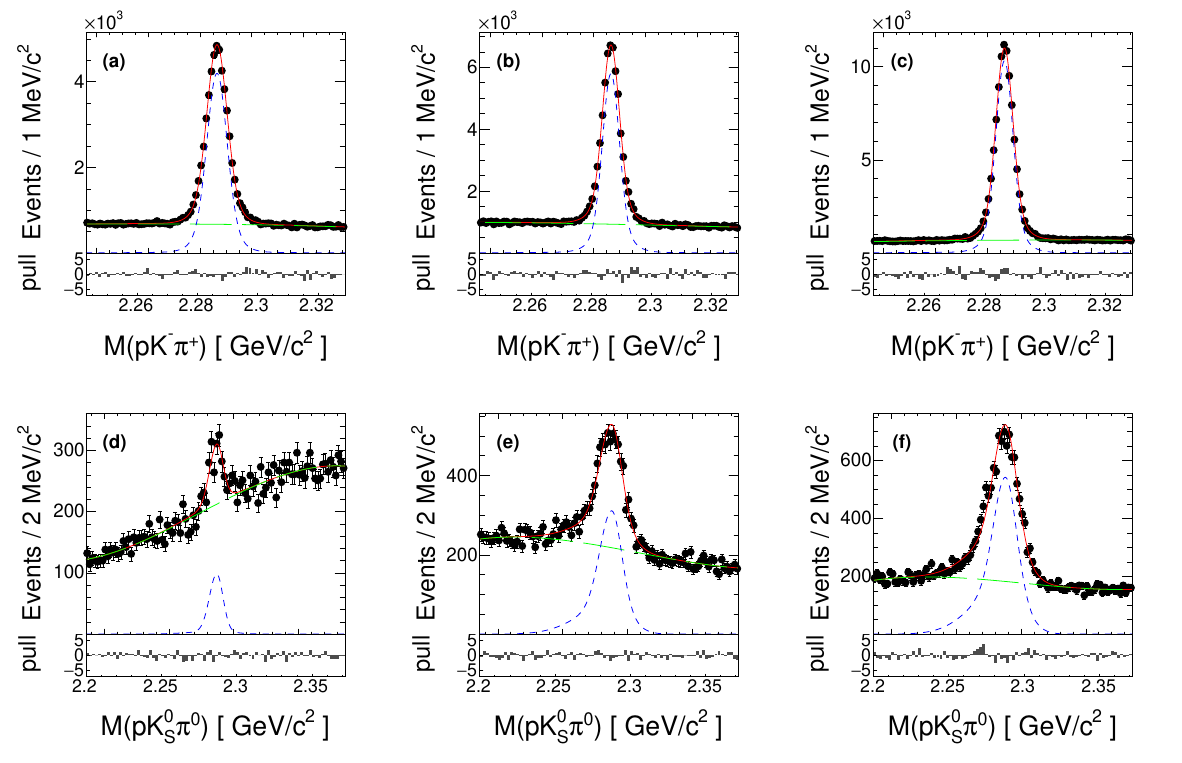}
\caption{Fit results~(red curves) in the sample Dalitz plot bins specified in Fig.~\ref{fig:yield} for~(a-c) $\Lccharged$ and~(d-f) $\Lcneutral$ decays. The dashed blue and long dashed green curves represent the signal and background, respectively.}
\label{fig:exampleFit}
\end{figure}
%%%%%

As the reconstruction efficiency varies across the phase space of
the Dalitz plot, as shown in Fig.~\ref{fig:efficiency},
we apply a bin-by-bin correction to estimate the efficiency-corrected yields of both decays.
To improve the resolution of the Dalitz plots, we fit the trajectories of the $\Lambda_c^+$ daughters to a common vertex and use a $\Lambda_c^+$ mass constraint.
The efficiency-corrected yield $y^{corr}$ calculated via
$y^{corr} = \sum\limits_{i}  {{y_i} \over \epsilon_{i}}$,
where $y_i$ and $\epsilon_{i}$ are 
the extracted yield and reconstruction efficiency of $i$-th bin in the Dalitz plots,
respectively.
In this correction, the Dalitz plots are divided into $5 \times 10$ bins
for both decays, as shown in Fig.~\ref{fig:yield}. 
Note that the bin width of the Dalitz plots is much larger than the resolution of the data, so the effect of bin migration on $y_{i}$ is negligible.

The yield for each bin is determined using the same functions employed
for the overall event fit.
However, all lineshape parameters, except for a common scaling factor applied to all
Gaussian widths, are fixed at the values obtained from corresponding signal MC samples.
This scaling factor corrects for the resolution difference between the data and the MC simulation.
The fitting results for typical bins are shown in Fig.~\ref{fig:exampleFit}.
We determine the efficiency-corrected yields as 
$(9.570 \pm 0.012) \times 10^{6}$ for $\Lccharged$
and $(2.221 \pm 0.015) \times 10^{6}$ for $\Lcneutral$, where the uncertainties 
are statistical only.
The bin-by-bin correction method enables the extraction of 
efficiency-corrected yields,
without requiring specific models for the production of intermediate states.

{\section{Branching Fraction}}
\label{section:BF}
The relative branching fraction is calculated using the following equation,

\begin{equation}
\hskip-0.4cm \frac{\mathcal{B}(\Lcneutral)}{\mathcal{B}(\Lccharged)} = 
\frac{y^{corr}(\Lcneutral)}{y^{corr}(\Lccharged) 
{\times \mathcal{B}(\pi^0 \to \gamma \gamma)
\times \mathcal{B}(\Kshort \to \pi^+\pi^-)}},
\label{eq:relativistic_eq}
\end{equation}
where we use $\mathcal{B}(\pi^0 \to \gamma \gamma) = (98.823 \pm 0.034) \%$
and  $\mathcal{B}(\Kshort \to \pi^+ \pi^-) = (69.20 \pm 0.05) \%$ from Ref.~\cite{PDG}.
By using Eq.~(\ref{eq:relativistic_eq}) and the efficiency-corrected yields,
the relative branching fraction is determined as follows:

\begin{equation}
\frac{\mathcal{B}(\Lcneutral)}{\mathcal{B}(\Lccharged)} = 0.339 \pm 0.002,
\label{eq:relativistic_value}
\end{equation}
where the uncertainty is statistical only.
\par

By assuming that the sum of the amplitudes,
$\sqrt{2}\mathcal{A}(p\bar{K}^{0}\pi^{0}) + \mathcal{A}(pK^{-}\pi^{+}) + \mathcal{A}(n\bar{K}^{0}\pi^{+})$, is zero as described above,
we can express the amplitudes in terms of two components,
$\mathcal{A}_{0}$ and $\mathcal{A}_{1}$,
corresponding to the isospin amplitudes
of the $I=0$ and $I=1$ states
of the $N\bar{K}$ system, respectively~\cite{Lu:2016ogy,Ablikim:2017}.
Defining a relative phase difference ($\delta$), between $\mathcal{A}_{0}$ and $\mathcal{A}_{1}$ as $\mathcal{A}_1/\mathcal{A}_{0} = |\mathcal{A}_1/\mathcal{A}_{0}|e^{i\delta}$, the relationship between the branching fractions and the isospin amplitudes is given by the following equations,

\begin{equation}
\mathcal{B}(\Lc \to p \bar{K}^0 \pi^0) = {1\over{2}} |\mathcal{A}_1|^2,
\label{eq:isospinamplitude1}
\end{equation}

\begin{equation}
\mathcal{B}(\Lccharged) = {1\over{2}} |\mathcal{A}_0|^2 + {1\over{4}} |\mathcal{A}_1|^2 -  {1\over{\sqrt{2}}} |\mathcal{A}_0||\mathcal{A}_1|\cos{\delta},
\label{eq:isospinamplitude2}
\end{equation}\\
 and
\begin{equation}
\mathcal{B}(\Lc \to n \bar{K}^0 \pi^+) = {1\over{2}} |\mathcal{A}_0|^2 + {1\over{4}} |\mathcal{A}_1|^2 +  {1\over{\sqrt{2}}} |\mathcal{A}_0||\mathcal{A}_1|\cos{\delta}.
\label{eq:isospinamplitude3}
\end{equation}

\par
With the measured value of
$\mathcal{B}(\Lcneutral) / \mathcal{B}(\Lccharged)$
and the world average $\mathcal{B}(\Lc \to n \bar{K}^0 \pi^+)/\mathcal{B}(\Lc \to p K^- \pi^+) = 0.581 \pm 0.084$~\cite{PDG},
$|\delta|$ was determined to be $1.842 \pm 0.069$, while the relative strength~($|\mathcal{A}_{1}|/|\mathcal{A}_{0}|$) was found to be $1.23 \pm 0.06$, where the uncertainty is the sum in quadrature of the statistical uncertainty and the uncertainty in $\mathcal{B}(\Lc \to n \bar{K}^0 \pi^+)/\mathcal{B}(\Lc \to p K^- \pi^+)$. 
The results show that the isospin amplitude $\mathcal{A}_{1}$ is
not significantly suppressed compared
to $\mathcal{A}_{0}$ in $\Lambda_{c}^{+}$ decays.
Here, we note that the calculation assumes isospin symmetry for the non-resonant contributions~\cite{Lu:2016ogy}.
\par

%%%%% 
\begin{figure}[!htb]
\centering
  \includegraphics[width=0.48\textwidth]{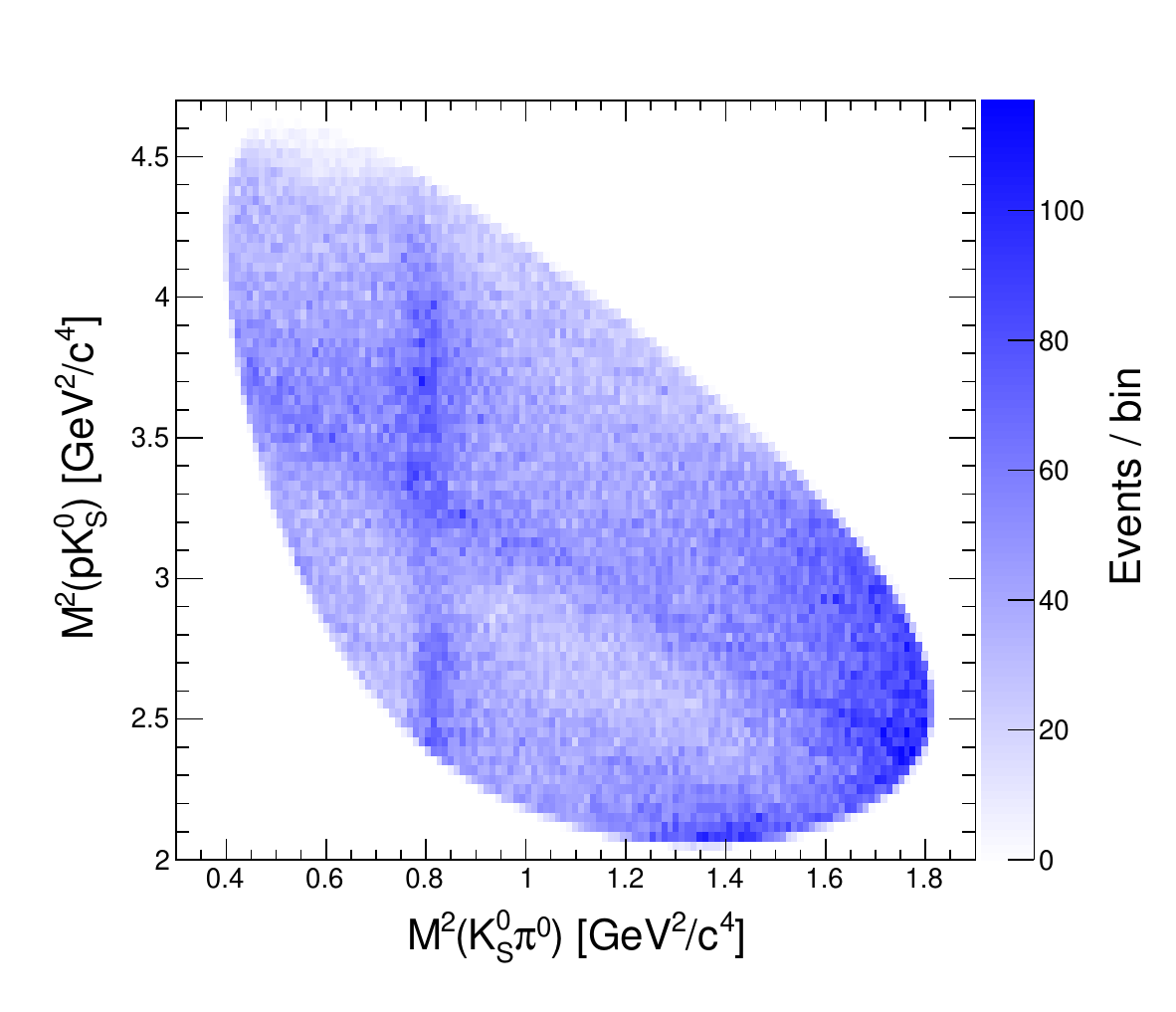}
  \includegraphics[width=0.48\textwidth]{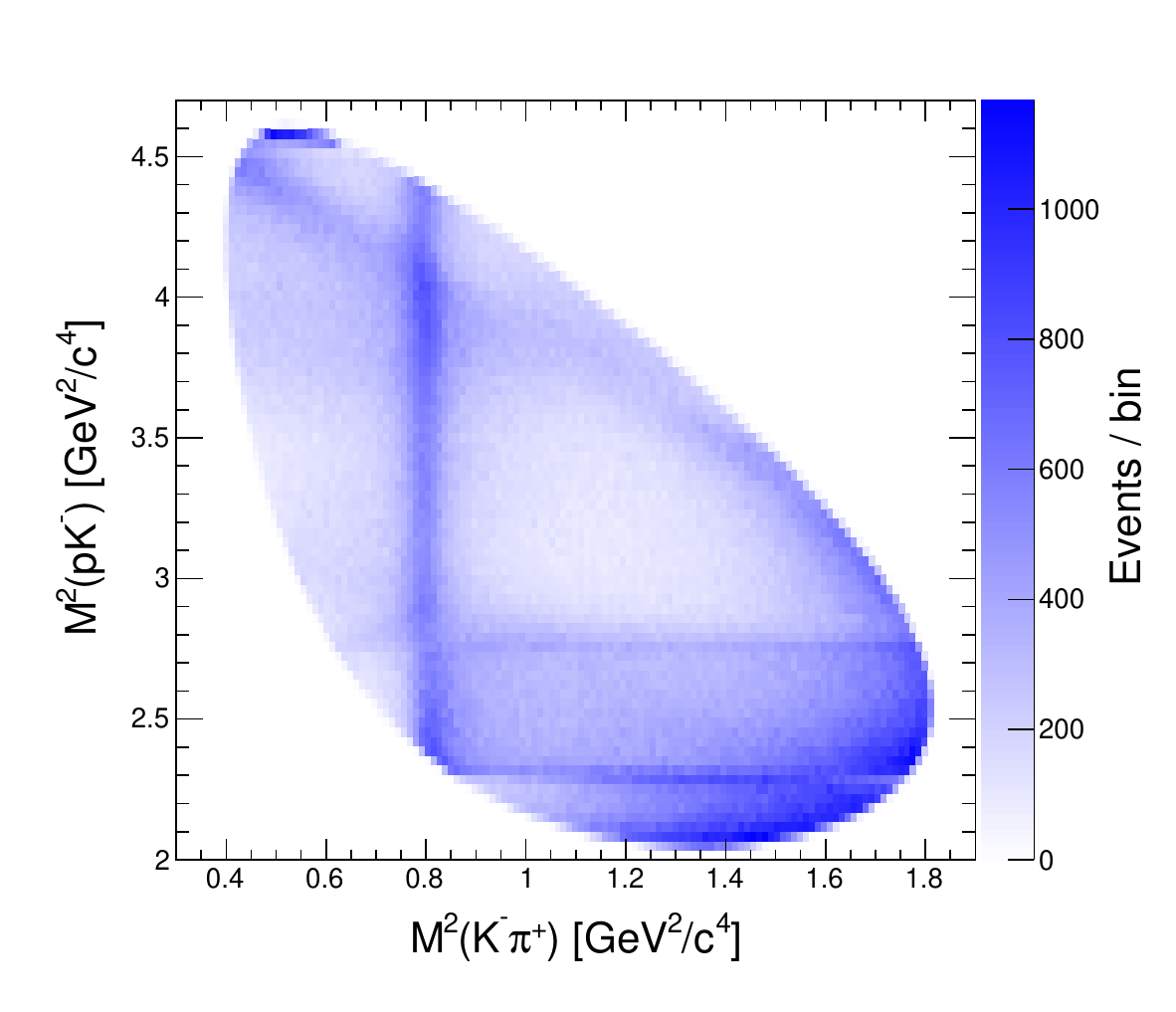}
\caption{
Dalitz plots of the $\Lcneutral$~(left) and $\Lccharged$~(right) channels
within the regions $2.263~\massGeV < M(pK_{S}^0\pi^0) <  2.306~\massGeV $
and $2.274~\massGeV < M(pK^-\pi^+) <  2.298~\massGeV $, respectively.
Both bin widths of $x$ and $y$ axes are 0.02~$\massGeVsq$.
Non-$\Lc$ background events shown in Fig.~\ref{fig:Lcmass} are included in the Dalitz plots.}
\label{fig:dalitz}
\end{figure}
%%%%%

%%%%%
\begin{figure}[!htb]
\centering
  \includegraphics[width=0.48\textwidth]{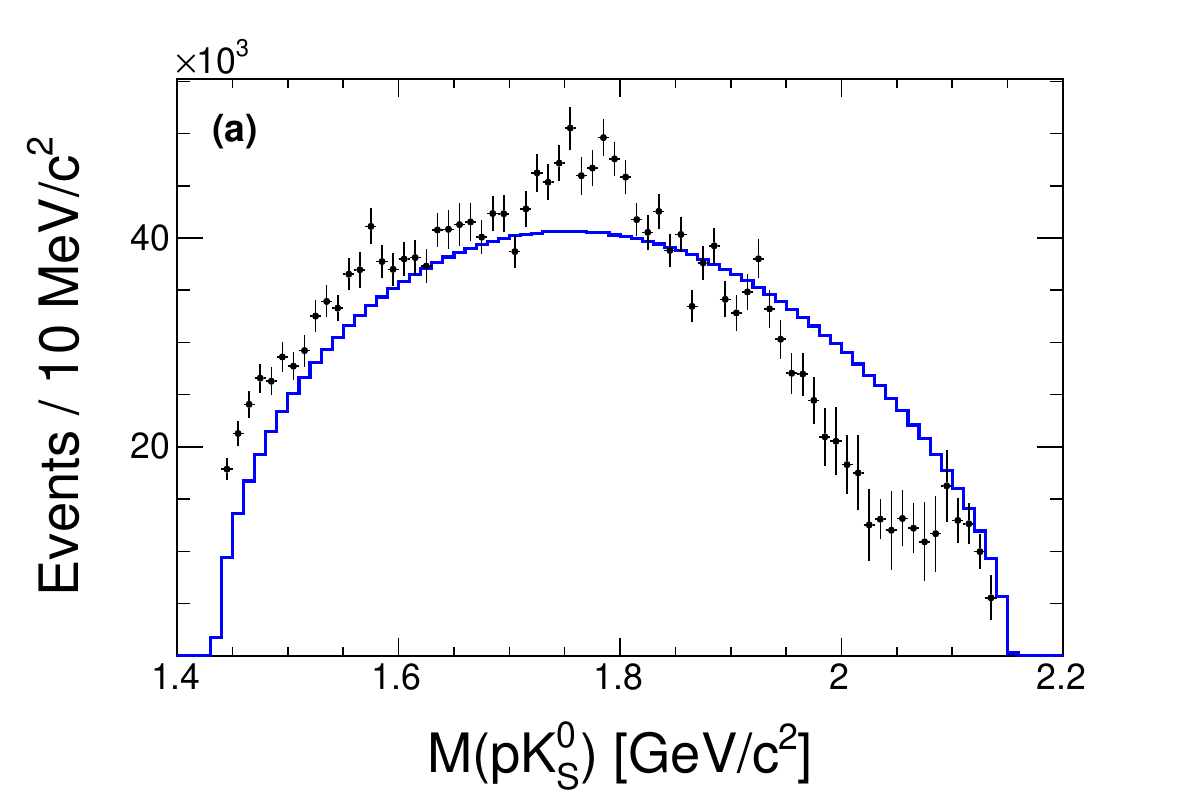}
  \includegraphics[width=0.48\textwidth]{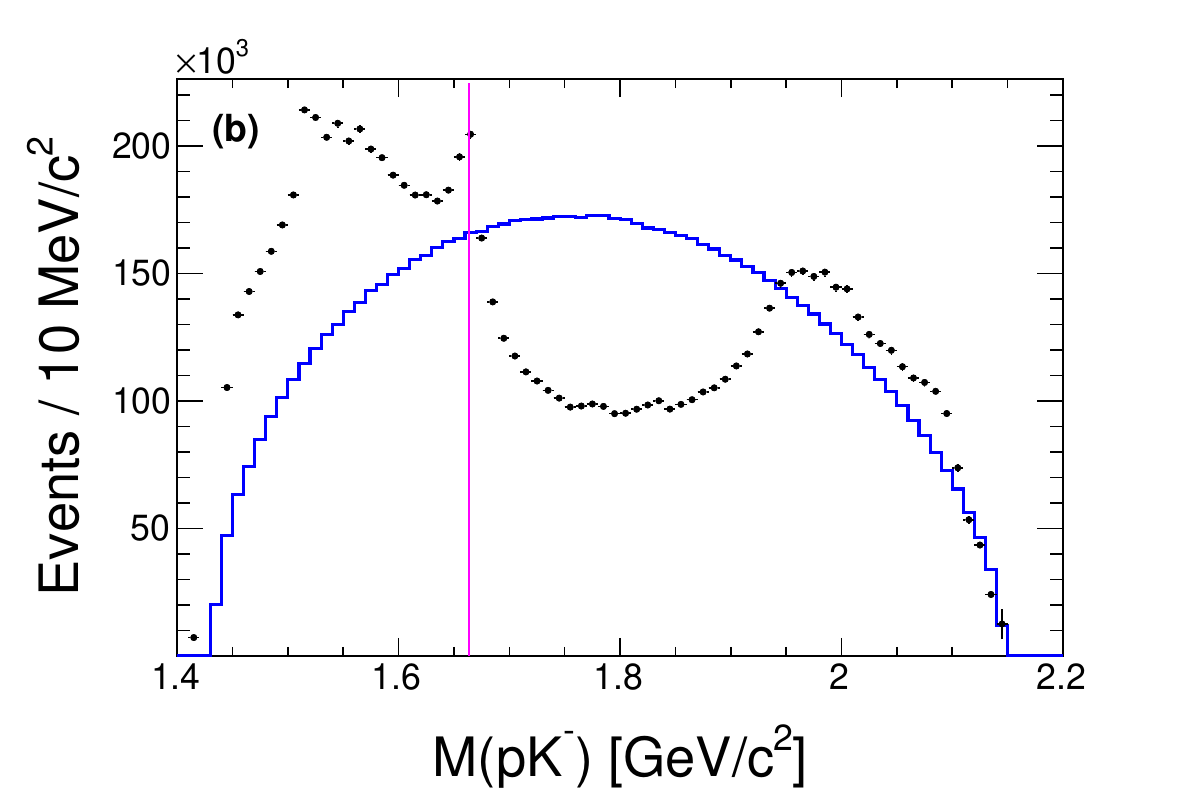}

  \includegraphics[width=0.48\textwidth]{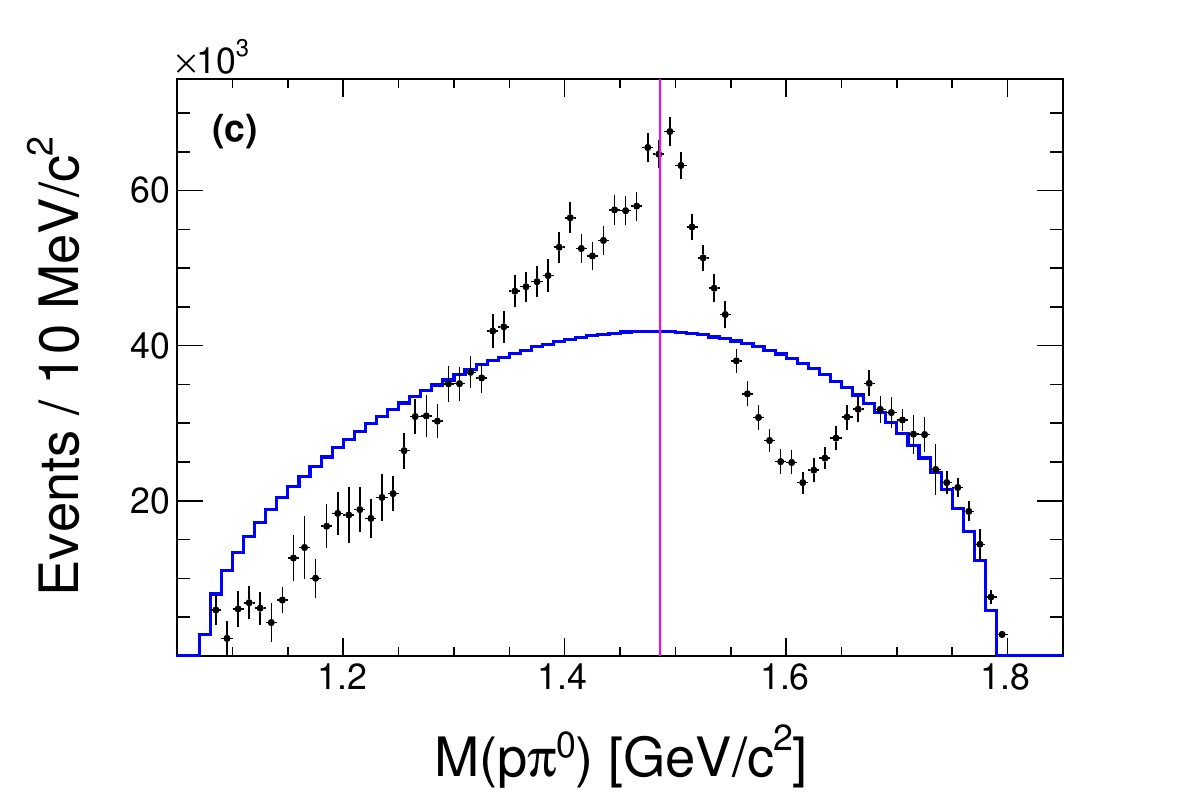}
  \includegraphics[width=0.48\textwidth]{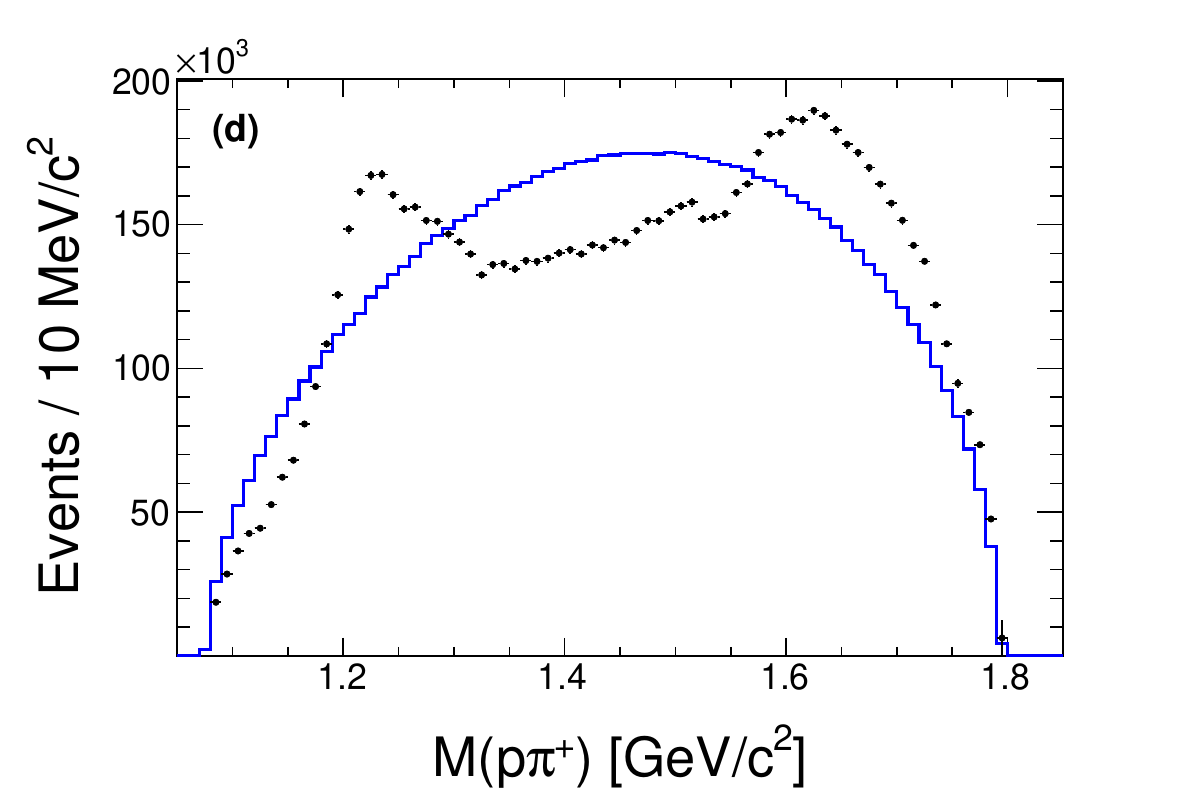}

  \includegraphics[width=0.48\textwidth]{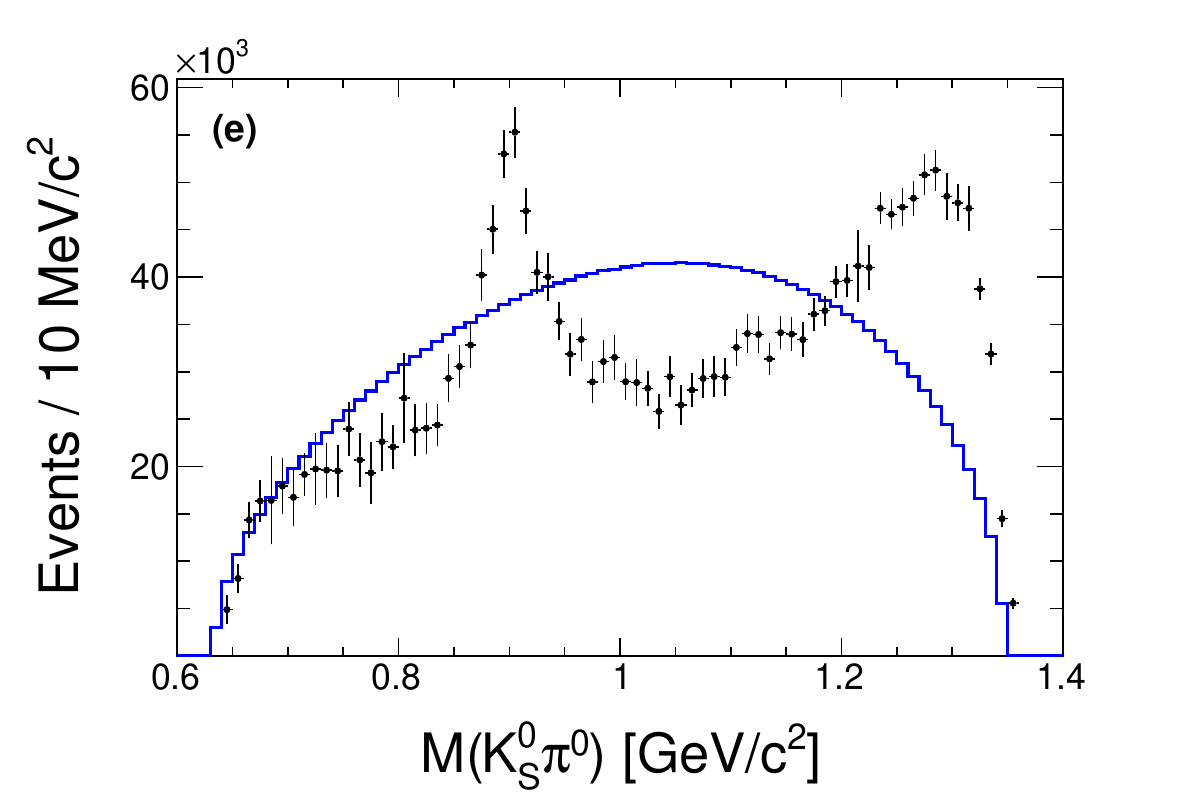}
  \includegraphics[width=0.48\textwidth]{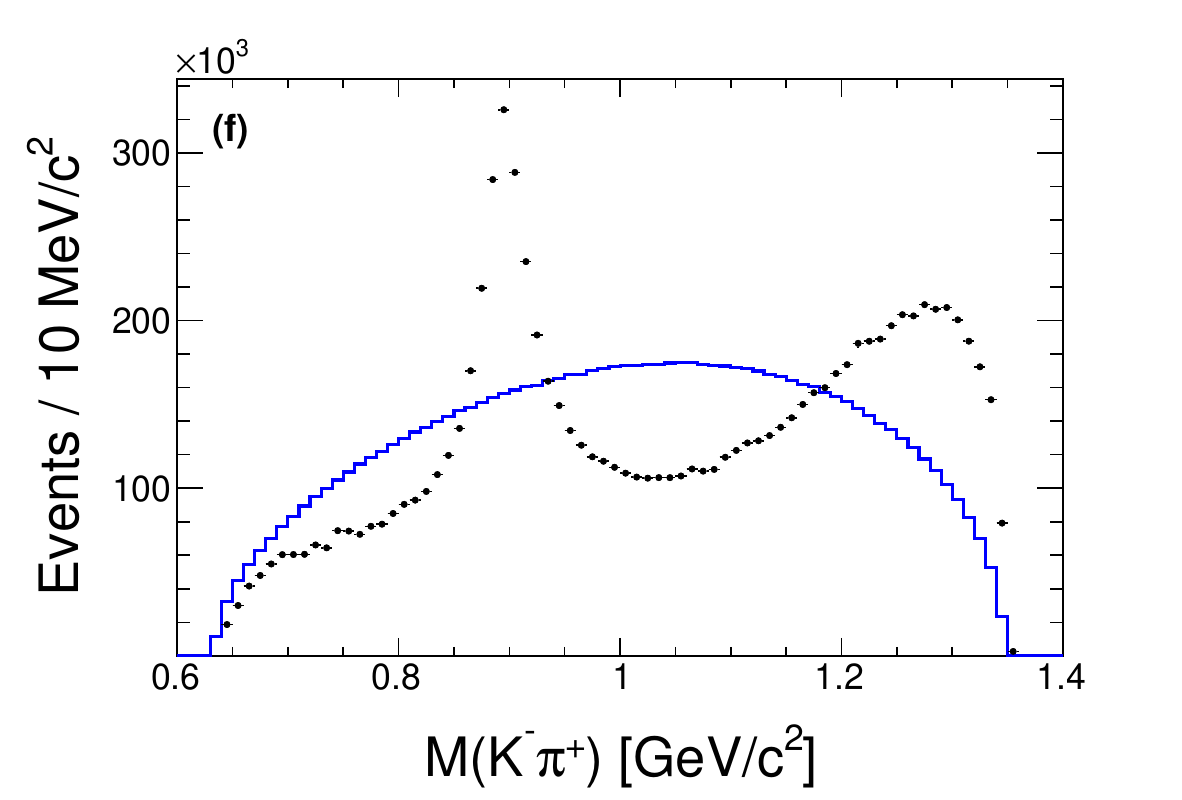}
  
\caption{Mass projection plots of $\Lambda_{c}^{+} \rightarrow p K^{0}_{S} \pi^{0}$~(left) and $\Lambda_{c}^{+} \rightarrow p K^{-} \pi^{+}$~(right) after background subtraction 
and efficiency correction. The projections of signal MC 
generated with a phase space model are superimposed in and shown as blue histograms. A strong enhancement near the $p\eta$ mass threshold is found in (c), 
and the $\Lambda\eta$ and $p\eta$ mass thresholds are marked in (b) and (c), respectively. 
A clear peaking structure at the $K^\ast$ resonance is seen in the $\Kshort\pi^0$ system (e).
}

\label{fig:massProjection}
\end{figure}

%%%%%

Figure~\ref{fig:dalitz} shows the Dalitz plot,
$M^{2}(K_{S}^{0}\pi^{0})$ vs. $M^{2}(pK_{S}^{0})$ for $\Lcneutral$ decays,
and several bands corresponding to intermediate resonances are observed in the plot.
We investigate the intermediate resonances by projecting the Dalitz plot onto
the one-dimensional distributions of $M(p\Kshort)$, $M(\Kshort\pi^0)$, and $M(p\pi^0)$. We apply efficiency corrections and then subtract non-$\Lambda_{c}^{+}$ background 
fitting the reconstructed $\Lambda_{c}^{+}$ 
mass distribution in each bin of plots with a similar method
described in Section \ref{section:BF}.

\par

In the $M(p\Kshort)$ distribution and the Dalitz plot of the $\Lcneutral$,
as shown in Fig.~\ref{fig:massProjection}(a) and the left of Fig.~\ref{fig:dalitz}, respectively,
$\Sigma^*$ hyperons are not as prominent. However, there are distinct peak structures that might be tentatively ascribed to $\Lambda(1520)$ and $\Lambda(1670)$ hyperons in the $M(pK^-)$ distribution, as shown in Fig.~\ref{fig:massProjection}(b).
This finding is in agreement with the expectation that $\Lambda^{*}$ hyperons are preferred over $\Sigma^{*}$ hyperons in the $\pi^{+}$ emission decays~\cite{Miyahara:2015}.
This could be attributed to their production dynamics being governed by color-suppressed factorizable process in the $\Lcneutral$ decay.

The peaking structure corresponding to $\Delta(1232)$
is much smaller in the $M(p\pi^{0})$ distribution of the $\Lcneutral$ sample~(Fig.~\ref{fig:massProjection}(c)) compared to the one in the $M(p\pi^+)$ distribution for $\Lccharged$ decays~(Fig.~\ref{fig:massProjection}(d)).
This suppression can be
attributed to the preference for the production
of the $\Delta^{++}K^{-}$ channel over the $\Delta^{+}\bar{K}^{0}$ channel,
as required by isospin symmetry.
Furthermore, a clear peaking structure near
the $p\eta$ mass threshold is evident in the $M(p\pi^{0})$ distribution
of $\Lcneutral$ decay.
This peak corresponds to the diagonal band
observed in the Dalitz plot to the left of Fig.~\ref{fig:dalitz}.
The same effect was observed in the $\Lambda_{c}^{+} \rightarrow pK_{S}^{0}\eta$ study in Ref.~\cite{Belle:2022pwd}.
The similarity of this effect and the $\Lambda\eta$ threshold cusp, which
was found to be amplified by the $\Lambda(1670)$ in the $pK^{-}$ system
as shown in Fig.~\ref{fig:massProjection}(b)~\cite{Belle:2022cbs,Belle:2020xku}, suggests that
the peak near the $p\eta$
threshold in the Fig.~\ref{fig:massProjection}(c) may also be attributed to a threshold cusp enhanced by the $N(1535)^{+}$.
\par

Both $\Lccharged$ and $\Lcneutral$ decays exhibit a peaking structure corresponding to the vector resonance $K^{*}(892)^{0}$, as shown in Fig.~\ref{fig:massProjection}(e) and~(f).
In the region where high-mass $K^{\ast}$ mesons are expected, a clear enhancement with respect to phase space is seen in both decay modes.
To further understand the role of isospin symmetry in the $\Lc$ decays and extend our understanding of intermediate states such as $\Lambda^{*}$, $\Sigma^{*}$, $\Delta^{*}$, and $N^{*}$ resonances, an amplitude analysis with the helicity formalism of these two channels is planned for the near future.

\section{Systematic Uncertainty}

The systematic uncertainties of branching fractions are listed in Table~\ref{tab:syst}. 
The $\Kshort$ reconstruction imposes a systematic uncertainty that has been estimated using a control sample of $D^{\ast\pm} \to D^0\pi^{\pm}$($D^0 \to \Kshort\pi^0$) events.
In the control sample study, the momentum-dependent $\Kshort$ reconstruction efficiency was compared between the data and MC samples.
In a similar way, the $\pi^0$ reconstruction uncertainty was calibrated by a study in Ref.~\cite{Belle:2014mfl} using $\tau^- \to \pi^- \pi^0 \nu_\tau$ events.
Here, the difference in efficiency for $M(\gamma\gamma)$ selection
between data and MC samples
is also added in quadrature to the systematic uncertainty.

The uncertainty due to the background model shown in Fig.~\ref{fig:exampleFit} is estimated
by changing it to a second-order
polynomial and a fourth-order polynomial.
We estimate the systematic uncertainty resulting from
the signal functions by performing one thousand fits, in which we vary the lineshape parameters fixed from the signal MC samples,
within their respective statistical uncertainties.
The systematic uncertainty is the standard deviation of the fit results.
The quadratic sum of the systematic uncertainties arising
from the background model and the signal function is
referred to as ``fit function" in Table~\ref{tab:syst}.

\begin{table}[!htb]
\begin{center}
\caption{Sources of systematic uncertainties 
for the relative branching fraction,~$\mathcal{B}(\Lambda^{+}_{c}\to pK^0_S\pi^0)/\mathcal{B}(\Lambda^{+}_{c}\to pK^-\pi^+)$.}
\begin{tabular}{cc}\\\hline \hline
%-------------------------------------------------------------
~~~~~~~~~~~~~~~~~~~~ Sources ~~~~~~~~~~~~~~~~~~~~ & ~~~~~~~~~~ Value~(\%) ~~~~~ \\\hline
$\Kshort$ reconstruction  & 1.57 \\
$\pi^0$ reconstruction & 1.54 \\
Fit function & 0.60 \\
MC statistics & 0.58 \\
Dalitz plot binning & 0.68 \\
PID of $K^-$ and $\pi^+$ & 0.34 \\
Tracking of $K^-$ and $\pi^+$ & 0.70 \\\hline
Total & 2.57 \\\hline \hline

\end{tabular}
\label{tab:syst}
\end{center}
\end{table}

We include the statistical uncertainty of the signal MC samples used in the efficiency corrections across the Dalitz plots as a systematic uncertainty.
We estimate the systematic uncertainties arising from the size of the Dalitz bins by modifying the Dalitz binning from the initial configuration of 5 $\times$ 10 to include the following configurations: 4 $\times$ 8, 4 $\times$ 10, 5 $\times$ 8, 5 $\times$ 12, 6 $\times$ 10, and 6 $\times$ 12.
The largest difference obtained is attributed to a corresponding systematic uncertainty. \par
The systematic uncertainty from $K^{-}$ and $\pi^{+}$ PIDs
in $\Lccharged$ decay
is calculated based on the
$D^{*+} \rightarrow D^{0}\pi^{+}(D^{0} \rightarrow K^{-}\pi^{+})$ control sample.
Similar to the $\Kshort$ reconstruction, the PID efficiency as a function of momentum and polar angle in the laboratory frame is compared between data and MC samples.
The systematic uncertainty attributed
to tracking is 0.35\% for each $K^{-}$ and $\pi^{+}$ track in $\Lccharged$ decay.

The uncertainties in the PDG values of $\mathcal{B}(\pi^0 \to \gamma\gamma)$ and $\mathcal{B}(\Kshort \to \pi^+ \pi^-)$ in Ref.~\cite{PDG} are negligible, so these contributions are not included in the systematic uncertainty.
Other systematic uncertainties cancel out for
the relative branching fraction measurements
due to the similar kinematic distributions of the final state particles from $\Lcneutral$ and $\Lccharged$ decays. 

%%%%%%%%%%%%%%%%%%%%%%%%%%%%%%%

\section{Summary}
We study the $\Lcneutral$ decay using the full Belle dataset of 980 $\rm fb^{-1}$ 
at or near the $\Upsilon(nS)$($n=1,2,3,4$, and 5) resonances. 
The branching fraction
of $\Lcneutral$ relative to $\Lccharged$ is determined as
\begin{equation}
  \frac{\mathcal{B}(\Lcneutral)}{\mathcal{B}(\Lccharged)} = 0.339\pm 0.002\pm 0.009,
\label{eq:relative_bf_summary}
\end{equation}
where the uncertainties are statistical and systematic, respectively. 
Using the PDG value of $\mathcal{B}(\Lccharged)=(6.24 \pm 0.28)\%$~in Ref.~\cite{PDG},
we obtain the following absolute branching fraction for $\Lcneutral$:
\begin{equation}
    \mathcal{B}(\Lcneutral) = (2.12\pm 0.01\pm 0.05 \pm 0.10) \%, 
\label{eq:absolute_bf_summary}
\end{equation}
where the uncertainties are statistical, systematic from this experiment and analysis,
and due to the uncertainty in $\mathcal{B}(\Lccharged)$, respectively.
The measured branching fraction is consistent with the previous measurement
by CLEO and has a fivefold improvement in precision~\cite{CLEO:1998dce}.
\par
Assuming isospin symmetry, we calculate
that the ratio of the isospin amplitudes for $I=1$ to $I=0$ in the $N\bar{K}$ system is determined to be $1.23 \pm 0.03 \pm 0.06$, and the relative phase difference is obtained to be $1.842\pm0.001\pm0.069$, where the first uncertainty denotes the combined statistical and experimental systematic uncertainty and the second uncertainty is from the ratio $\mathcal{B}(\Lc \to n \bar{K}^0 \pi^+)/\mathcal{B}(\Lc \to p K^- \pi^+)$.
These values are consistent with previous results \cite{Ablikim:2017}.
However, we do not find a strong enhancement due to $\Sigma^{*}$ resonances in the $M(pK^{0}_{S})$ distribution of $\Lcneutral$ decay.
These results indicate that factors beyond isospin symmetry,
such as resonant contributions, cannot be neglected.

In addition, we observe a clear peaking structure in the $p\pi^0$ system near the $p\eta$ threshold, which may be attributed to a threshold cusp enhanced by $N(1535)^{+}$.
Further amplitude analysis is required to understand the contributions of intermediate resonances such as $K^*$, $\Lambda^{*}$, $\Sigma^{*}$, $\Delta^*$, and $N^{*}$ resonances, as well as to estimate the non-resonant contribution.
Such a comprehensive approach will lead to stringent tests of isospin symmetry by comparing the partial branching ratios between $\Lccharged$ and $\Lcneutral$ decays. This approach could also contribute to a better understanding of non-factorizable processes in the non-leptonic decay of charmed baryons.

%\clearpage
\section*{Acknowledgments}
%\input{acknowledgements}
% Policy from October 20, 2022
This work, based on data collected using the Belle II detector, which was built and commissioned prior to March 2019,
and data collected using the Belle detector, which was operated until June 2010,
was supported by
%Armenia
Higher Education and Science Committee of the Republic of Armenia Grant No.~23LCG-1C011;
%Australia
Australian Research Council and Research Grants
No.~DP200101792, % Jackson
No.~DP210101900, % Urquijo
No.~DP210102831, % Sevior
No.~DE220100462, % Hsu
No.~LE210100098, % Infrastructure
and
No.~LE230100085; % Infrastructure
%Austria
Austrian Federal Ministry of Education, Science and Research,
Austrian Science Fund (FWF) Grants
DOI:~10.55776/P34529,
DOI:~10.55776/J4731,
DOI:~10.55776/J4625,
DOI:~10.55776/M3153,
and
DOI:~10.55776/PAT1836324,
and
Horizon 2020 ERC Starting Grant No.~947006 ``InterLeptons'';
%Canada
Natural Sciences and Engineering Research Council of Canada, Compute Canada and CANARIE;
%China
National Key R\&D Program of China under Contract No.~2022YFA1601903,
National Natural Science Foundation of China and Research Grants
No.~11575017,
No.~11761141009,
No.~11705209,
No.~11975076,
No.~12135005,
No.~12150004,
No.~12161141008,
No.~12475093,
and
No.~12175041,
and Shandong Provincial Natural Science Foundation Project~ZR2022JQ02;
%Czech Republic
the Czech Science Foundation Grant No.~22-18469S 
and
Charles University Grant Agency project No.~246122;
%EU
European Research Council, Seventh Framework PIEF-GA-2013-622527,
Horizon 2020 ERC-Advanced Grants No.~267104 and No.~884719,
Horizon 2020 ERC-Consolidator Grant No.~819127,
Horizon 2020 Marie Sklodowska-Curie Grant Agreement No.~700525 ``NIOBE''
and
No.~101026516,
and
Horizon 2020 Marie Sklodowska-Curie RISE project JENNIFER2 Grant Agreement No.~822070 (European grants);
%France
L'Institut National de Physique Nucl\'{e}aire et de Physique des Particules (IN2P3) du CNRS
and
L'Agence Nationale de la Recherche (ANR) under Grant No.~ANR-21-CE31-0009 (France);
%Germany
BMBF, DFG, HGF, MPG, and AvH Foundation (Germany);
%India
Department of Atomic Energy under Project Identification No.~RTI 4002,
Department of Science and Technology,
and
UPES SEED funding programs
No.~UPES/R\&D-SEED-INFRA/17052023/01 and
No.~UPES/R\&D-SOE/20062022/06 (India);
%Israel
Israel Science Foundation Grant No.~2476/17,
U.S.-Israel Binational Science Foundation Grant No.~2016113, and
Israel Ministry of Science Grant No.~3-16543;
%Italy
Istituto Nazionale di Fisica Nucleare and the Research Grants BELLE2,
and
the ICSC – Centro Nazionale di Ricerca in High Performance Computing, Big Data and Quantum Computing, funded by European Union – NextGenerationEU;
%Japan
Japan Society for the Promotion of Science, Grant-in-Aid for Scientific Research Grants
No.~16H03968,
No.~16H03993,
No.~16H06492,
No.~16K05323,
No.~17H01133,
No.~17H05405,
No.~18K03621,
No.~18H03710,
No.~18H05226,
No.~19H00682, % Niigata
No.~20H05850,
No.~20H05858,
No.~22H00144,
No.~22K14056,
No.~22K21347,
No.~23H05433,
No.~26220706,
and
No.~26400255,
%the National Institute of Informatics, and Science Information NETwork 5 (SINET5), 
and
the Ministry of Education, Culture, Sports, Science, and Technology (MEXT) of Japan;  
%Korea
National Research Foundation (NRF) of Korea Grants
No.~2016R1-D1A1B-02012900,
No.~2018R1-A5A-1025563, % KIM, YANG, AHN
No.~2018R1-A6A1A-06024970,
No.~2020R1-A3B-2079993, % KIM, YANG, AHN 
No.~2021R1-A6A1A-03043957,
No.~2021R1-C1C-2012925, % KIM, YANG, AHN 
No.~2021R1-F1A-1060423,
No.~2021R1-F1A-1064008,
No.~2022R1-A2C-1003993,
No.~2022R1-A2C-1092335,
No.~RS-2023-00208693,
No.~RS-2024-00354342
and
No.~RS-2022-00197659,
Radiation Science Research Institute,
Foreign Large-Size Research Facility Application Supporting project,
the Global Science Experimental Data Hub Center, the Korea Institute of
Science and Technology Information (K24L2M1C4)
and
KREONET/GLORIAD;
%Malaysia
Universiti Malaya RU grant, Akademi Sains Malaysia, and Ministry of Education Malaysia;
%Mexico
% CINVESTAV-IPN, UNAM, UAS, BUAP and CONACYT are funded under
Frontiers of Science Program Contracts
No.~FOINS-296,
No.~CB-221329,
No.~CB-236394,
No.~CB-254409,
and
No.~CB-180023, and SEP-CINVESTAV Research Grant No.~237 (Mexico);
%Poland
the Polish Ministry of Science and Higher Education and the National Science Center;
%Russia
the Ministry of Science and Higher Education of the Russian Federation
and
the HSE University Basic Research Program, Moscow;
%Saudi Arabia
University of Tabuk Research Grants
No.~S-0256-1438 and No.~S-0280-1439 (Saudi Arabia), and
Researchers Supporting Project number (RSPD2025R873), King Saud University, Riyadh,
Saudi Arabia;
%Slovenia
Slovenian Research Agency and Research Grants
No.~J1-9124
and
No.~P1-0135;
%Spain
Ikerbasque, Basque Foundation for Science,
the State Agency for Research of the Spanish Ministry of Science and Innovation through Grant No. PID2022-136510NB-C33,
Agencia Estatal de Investigacion, Spain
Grant No.~RYC2020-029875-I
and
Generalitat Valenciana, Spain
Grant No.~CIDEGENT/2018/020;
%Swiss (Belle 1)
the Swiss National Science Foundation;
%Sweden
The Knut and Alice Wallenberg Foundation (Sweden), Contracts No.~2021.0174 and No.~2021.0299;
%Taiwan
National Science and Technology Council,
and
Ministry of Education (Taiwan);
%Thailand
Thailand Center of Excellence in Physics;
%Turkey
TUBITAK ULAKBIM (Turkey);
%Ukraine
National Research Foundation of Ukraine, Project No.~2020.02/0257,
and
Ministry of Education and Science of Ukraine;
%USA
the U.S. National Science Foundation and Research Grants
No.~PHY-1913789 % Indiana CEEM
and
No.~PHY-2111604, % Luther
and the U.S. Department of Energy and Research Awards
No.~DE-AC06-76RLO1830, % PNNL
No.~DE-SC0007983, % Wayne State
No.~DE-SC0009824, % Florida
No.~DE-SC0009973, % VPI
No.~DE-SC0010007, % Duke
No.~DE-SC0010073, % South Carolina
No.~DE-SC0010118, % Carnegie Mellon
No.~DE-SC0010504, % Hawaii
No.~DE-SC0011784, % Cincinnati
No.~DE-SC0012704, % BNL
No.~DE-SC0019230, % Duke
No.~DE-SC0021274, % Mississippi
No.~DE-SC0021616, % Mississippi
No.~DE-SC0022350, % Louisville
No.~DE-SC0023470; % South Alabama
%last group
and
%Vietnam
the Vietnam Academy of Science and Technology (VAST) under Grants
No.~NVCC.05.12/22-23
and
No.~DL0000.02/24-25.

% Policy from October 20, 2022
These acknowledgements are not to be interpreted as an endorsement of any statement made
by any of our institutes, funding agencies, governments, or their representatives.

We thank the SuperKEKB team for delivering high-luminosity collisions;
the KEK cryogenics group for the efficient operation of the detector solenoid magnet and IBBelle on site;
the KEK Computer Research Center for on-site computing support; the NII for SINET6 network support;
and the raw-data centers hosted by BNL, DESY, GridKa, IN2P3, INFN, 
PNNL/EMSL, 
and the University of Victoria.
%Y. J. Kim, S. B. Yang, and J. K. Ahn acknowledge the support provided by NRF Grant No. 2018R1A5A1025563, No. 2020R1A3B2079993, and No. 2021R1C1C2012925.


\begin{thebibliography}{99}
\bibitem{conjugate}
Unless otherwise noted, charge-conjugate modes are always implied throughout this paper.
\end{thebibliography}
\end{document}